\documentclass{aa501}
\usepackage{graphicx}

\def\m{$\cal M$}
\def\ml{\m/L}

\def\lya{Ly$\alpha$~}
\def\ha{H$\alpha$~}
\def\hb{H$\beta$~}
\def\xha{H$\alpha$}
\def\xhb{H$\beta$}
\def\hahb{\xha/\hb}
\def\wha{W$_{H\alpha}$~}

\def\xwha{W$_{H\alpha}$}

\def\sm{$\cal M_\odot$~}
\def\sma{$\cal M_\odot$}
\def\ma{$\cal M$}

\begin{document}

\title{Massive (?) starburst hosts of blue compact galaxies (BCGs)}
\subtitle{Optical/near-IR observations of 4 BCGs and their
companions\thanks{Based
on observations collected at the European
Southern Observatory, La Silla, Chile.}}

\author{ Nils Bergvall\inst{1}
\and 
G\"oran \"Ostlin\inst{2}}

\offprints{N. Bergvall}

\institute{Dept. of Astronomy and Space Physics, Box 515, S-75120 Uppsala, 
Sweden 
\\
\email{nils.bergvall@astro.uu.se}
\and
Stockholm Observatory, SCFAB, SE-106 91 Stockholm, Sweden \\
\email{ostlin@astro.su.se}
}

\date{Received 29 November 2001; accepted 15 May 2002}

\abstract{
We present optical spectroscopy and deep optical/near-IR photometry of 4 
luminous metal-poor blue compact galaxies (BCGs) and two of their companions. 
With the aid of spectral evolutionary models (SEMs) and structural parameters 
derived from the surface photometry we discuss the properties of the central 
starbursts and the halo populations of the galaxies. Special attention is paid 
to the effects of dust, chemical inhomogeneities and contamination of nebular 
emission to the halo light. The optical/near-IR colour index profiles show a 
sharp distinction between the starburst and the host. The hosts have luminosity 
profiles characteristic of massive ellipticals and remarkably red colours, 
typical of a relatively {\it metal-rich} stellar population of {\it old age}. 
These properties are in conflict with the relatively low luminosities. The 
situation can best be explained if the hosts have an unusually large amount 
of dark matter that can hinder the outflow of metals from the system.
The 
indicated difference in metallicity between the halo and the young starburst 
disproves the recurrent burst scenario and supports different origins of the two 
populations. We conclude that these BCGs are undergoing mergers between early 
type galaxies/thick disks and gas-rich galaxies or intergalactic HI 
clouds, in many respects reminiscent of a retarded formation of massive 
ellipticals. 
}

\authorrunning{Bergvall and \"Ostlin}

\titlerunning{The halo of four BCGs}

\maketitle

\section{Introduction}
\subsection{General comments}

Blue compact galaxies (BCGs), sometimes called HII galaxies, are characterized
by globally active star formation, low chemical abundances (e.g. Searle \& 
Sargent
\cite{searle}, Marconi et al. \cite{marconi}, Kunth \& \"Ostlin \cite{kunth2},
Masegosa et al. \cite{pepa}) and relatively high HI masses (Gordon and Gottesman
\cite{gordon}, Thuan and Martin \cite{thuan2}, Staveley-Smith et al. 
\cite{staveley}). These are properties normally associated
with young galaxies and indeed there have been claims from time to time (Searle
\& Sargent \cite{searle}, Thuan \& Izotov \cite{thuan1}) that some BCGs may
be truly young systems. But the most important aspect today is that BCGs
constitute an important link to the high redshift universe and the early
epoch of galaxy formation.

Observations (e.g. Lilly et al. \cite{lilly}, Le F\`evre et al. \cite{lefevre})
show that mergers between galaxies are of major importance for
the buildup of galaxies at high redshifts. The BCGs and their progenitors 
(HI clouds, LSB galaxies or other gas-rich dwarfs) that contribute fuel
for the starburst may therefore be regarded as the local analogues
of the distant subunits participating in the early merger processes. While 
starbursts of massive galaxies are rare, those induced by mergers of galaxies
of intermediate mass may produce a major fraction of the metals observed in
the intergalactic medium (IGM) during the buildup processes at high redshifts.
A necessary requirement is that the energy input from the exploding supernovae
is sufficiently large to overcome the gravitation. This requires masses of the
galaxies not much in excess of typical luminous BCGs, making this type of
galaxy interesting also from this point of view. 

For a better understanding of the star formation processes we want to
disentangle the different
phases involved in the formation of a BCG by identifying stellar populations
and galaxy morphologies of components from
different epochs. We also want to know under what conditions a starburst can be
triggered. Are the conditions set by external conditions, like the orbital
properties of infalling clouds, or is a specific type of galaxy needed to host
the burst?

\subsection{Ages of BCGs and the youth hypothesis}

From studies of the central starburst region of BCGs, it is very difficult
to rule out that the galaxy is young since the starburst population easily 
outshines even a relatively massive population of old stars. Therefore, 
some investigations have focused on the halos  of BCGs, where the starburst
influence is assumed to be milder. 
Most BCGs show a regular halo (e.g. Loose and Thuan \cite{loose}). 
Already this fact is a strong
argument in favour of a fairly old age because the relaxation
time of a system of stars of a typical size of a few kpc is larger than a few
times 10$^8$ yr. But a regular halo can be formed on a shorter time scale if
the medium is viscous. The light may be
due to nebular emission, a "Str\"omgrensphere" formed by the
starburst in the centre. We will elaborate on this a bit further in
section 6. If so, the stellar component whose formation may
be much delayed relative to the gaseous disk of a protogalaxy,
may be quite young.

One of the most debated young galaxy candidates is
IZw18, which  belongs to the blue compact dwarfs (BCDs).
These are galaxies in the mass range 10$^7$ - 10$^9$ \sma, 
having relatively low luminosities ($M_B >-17$).  IZw18 
has unique properties in this category, in particular 
its record-low oxygen abundance of approximately 1/50 of the solar value
(e.g. Alloin et al. \cite{alloin}; Izotov and Thuan \cite{izotov}).
There are however reasons to suspect that winds from massive stars 
and supernovae in low-mass galaxies may  expel the gas and lower 
the chemical abundances in which case 
the youth signature may not be more than an artifact (e.g. McLow and 
Ferarra \cite{maclow}, Lequeux et al \cite{lequeux}, Martin \cite{martin}). 
It is also a bit difficult to cope the relatively low \m (HI)/M$_B$ 
$\sim$ 1 in IZw18 (Thuan and Martin 1981) with a recently born
galaxy. As we show below (Fig.~\ref{jhkgen}), already a "simple" set of
data such as the integrated near-IR colours indicate that this is
an old galaxy. Recent investigations of deep colour-magnitude 
diagrams of IZw18, in the optical by Aloisi et al. (1999) and in the
near IR by \"Ostlin (2000), show that the galaxy contains a population 
of evolved red stars, indicating an age in excess of 1 Gyr. 
Among other young galaxy candidates discussed in the literature 
(Bergvall \& J\"ors\"ater \cite{bergvall2}) is one of the targets 
in this project, ESO 400-G43, but as we will show below this galaxy 
also seems to contain old stars. Another debated young galaxy candidate is
SBS0335-052 (Thuan \& Izotov \cite{thuan1}; \"Ostlin \& Kunth \cite{ok}).

Near-IR photometry has proven to be quite powerful in deriving information 
about the star formation history in galaxies that are dominated by starbursts
and seems to support the idea that BCGs are old (e.g. Bergvall et al. 
\cite{bergvall8}, Doublier et al. \cite{doublier2}). 
Then the interesting question remains what kind of
galaxy or galaxies the precursor(s) and the successors could be.
Several different scenarios for the ignition of the burst and the
types of galaxies involved have been discussed in the literature
(e.g. Searle \& Sargent \cite{searle}, Thuan \& Seitzer \cite{thuan3}, 
Thuan \&
Martin \cite{thuan2}, Staveley-Smith et al.
\cite{staveley}, Taylor et al. \cite{taylor}, Telles \& Terlevich
\cite{telles},
Papaderos et al. \cite{papaderos}). One should be careful not to consider
all BCDs as starburst galaxies, as is the commonly accepted
view. In fact Sage et al. (\cite{sage}) find that most BCDs are not
more efficient in converting gas into stars than are normal
spiral galaxies. They appear to be bursting only when
compared to other gas-rich dwarfs where the normal star
formation rate (SFR) is lower than in massive gas-rich
galaxies. It is important to keep in mind that the class of BCGs is
quite heterogeneous (Kunth et al. \cite{kunth}; Kunth \& \"Ostlin \cite{kunth2}) 
and may include
objects with different histories. Some galaxies called BCGs by some
groups are definitely not BCGs since they easily fit into the extended
Hubble classification towards late type disks or normal irregulars
(Sandage \& Binggeli \cite{sandage}).

In this work we are dealing with galaxies that have
extremely high {\it star formation efficiencies} (SFE). With SFE we mean the
timescale of gas consumption in a closed box scenario, and a short timescale is 
equal 
to a high SFE. From our spectral
 evolutionary models
(SEMs), assuming a Salpeter mass function (\cite{salpeter}) and a mass range of 
0.1-120 \sm we
find a SFR of $\sim$ 10 \sm yr$^{-1}$ for a burst with M$_B$ between -19 and
-20. Assuming a gas mass of 10$^9$ \sm, which is typical for our sample,  we 
thus obtain a maximum lifetime of
the burst of not more than a few times 10$^8$ yr, allowing for a modest fading 
of 
the burst. Thus, if no fresh gas is supplied the starburst will be a
transient phenomenon during which the properties of the galaxy will change
from what it was before, allowing a morphological metamorphosis to take
place. Based on photometric properties, globular cluster properties, HI
masses, kinematics and chemical abundances (e.g. Bergvall et al. 
\cite{bergvall5},
\"Ostlin et al. \cite{goran1}, \cite{goran3}) we have claimed that
a large fraction of luminous BCGs may form from mergers involving massive gas
rich low surface brightness galaxies or gas clouds and possibly early type
dwarfs. More support in favour of this will be presented below as we discuss 
the properties of 4 such massive BCGs. The relationships between H{\sc i}, 
chemical
abundances and  photometric properties will be discussed in a paper now in
preparation.

\subsection{Outflows of enriched gas}

As noted below, we have observational support for bipolar outflows
 from the BCGs
  discussed here. An interesting question to ask is under
  what conditions and to what extent the metals in these outflows can reach the
  escape velocity and be expelled into the ambient intergalactic medium (IGM) 
and how much of the metals observed
in \lya systems and the hot intracluster gas in rich galaxy clusters that can 
be explained this way. Outflows will also open channels in the gaseous
halos, making it easier for Ly-continuum photons to escape. Starburst dwarfs may 
therefore contribute significantly to the
reionisation of the IGM after recombination. 

McLow and Ferarra (\cite{maclow}) argue that galaxies with masses exceeding
5 10$^7$ \sm (less than a few \% of the masses of our targets) will retain 
most of their gas indefinitely, while a large fraction of the metals may
be lost. If a bottom-up scenario is relevant for early galaxy formation it could 
be argued that most of the pollution of the IGM in the early days were caused
by outflows from starburst dwarfs, similar to the local BCGs. These important
topics deserve a thorough investigation   but is not the prime goal of this
paper. Still, as we illustrate below, we have to consider the possibility that
the outflows influence the properties of the halo through polluting the light 
with  
emission from ionised gas of low temperature.

\subsection{Near-IR properties of BCGs in general}

\begin{figure}
\includegraphics[width=9cm]{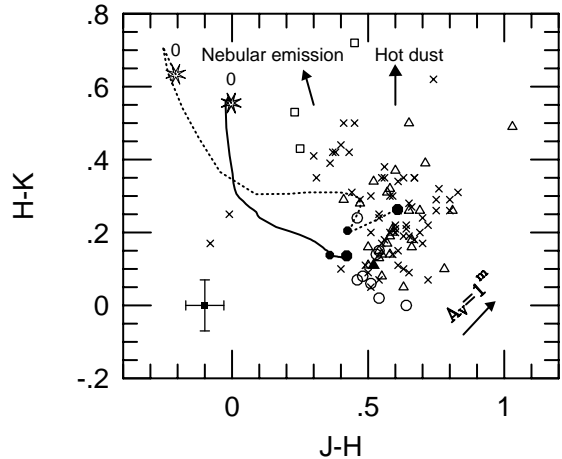}
\caption[]{The J-H/H-K two colour diagram of BCGs (crosses), luminous irregulars 
(triangles), HII regions (squares) and dE:s (circles) obtained from Bergvall
\& Olofsson (\cite{bergvall1}), Hunter \& Gallagher (\cite{hunter}), James 
(\cite{james1}) and Thuan (\cite{thuan0}). The predicted evolution (Zackrisson 
et al. \cite{erik}) of a star forming galaxy with a Salpeter mass function, an 
e-folding star formation decay rate of 14 Gyr and two different metallicities 
(20\% solar, solid line; twice solar, dotted line) are also displayed. The 
evolution 
starts at the star marked 0 and ends at the large black dot, corresponding to an 
age 
of 14.5 Gyr. The position at 1 Gyr is marked with a smaller black dot. The 
effect 
of
dust reddening and emission are indicated by arrows but note that the nebular
emission has been included in the model. The filled triangle at J-H=0.5,
H-K=0.1 is IZw18. Typical mean errors are indicated on the lower left side.}  
\label{jhkgen}   
\end{figure}

Before entering into the results of this paper, it is enlightening 
to have a look at the integrated near-IR properties of BCGs in general. 
In Fig.~\ref{jhkgen} we show the $J-H$ vs $H-K$ diagram of BCGs, 
dIs and dEs with available photometry from the literature. In the
diagram we have also indicated the evolution of a star 
forming galaxy with a Salpeter IMF and an exponentially decaying  
SFR (e-folding time 14 Gyr). Two tracks are shown, representing different 
metallicities
($0.2 Z_\odot$ and $2 Z_\odot$). The galaxy positions are quite 
dispersed in the diagram. As mentioned above,  BCGs in general 
is a very heterogeneous group of galaxies, composed of subgroups with 
different evolutionary histories.  Despite the spread in 
Fig.~\ref{jhkgen} it is obvious that most, if not all, galaxies  
contain old stars. 

Some spread may be due to varying data quality, and 
differences in metallicity and extinction.
However, even when these effects are taken into account, it seems 
as if the observed distribution is shifted towards slightly redder 
$J-H$. The difference can be reduced with 0.1 magnitudes if we assume 
that the star formation rate declines faster so that the influence 
from young stars becomes less important.   
In the cases where the galaxy contains an old population of
stars and the starburst is not so strong we would expect that the old
stars dominate in the infrared. This is indeed where most of
the BCGs are found in the diagram. Only a few BCGs with strong
starbursts are located along the evolutionary track with low ages.

\begin{table*}
\caption[ ]{\bf Log of optical imagery}
\begin{flushleft}
\begin{tabular} {llllllll}
\noalign{\smallskip}
\hline
\noalign{\smallskip}
ESO id. & R.A.$_{1950}$  & Decl.$_{1950}$  & Year & Telescope &
Filters and integration times (minutes) & Seeing  \cr
& & & & & & (arcsec) \\
\noalign{\smallskip}
\hline
\noalign{\smallskip}
338-04 (Tol 1924-416)& 19 24 29.0  & -41 40 36  & 1995 & NTT  & \ha (10), 
H$\alpha$-continuum (33) & 1 \cr
&&& 1997 & NTT  & B(5), V(15), R(15), I(30) & 1 \cr
338-04b & 19 24 03 & -41 45 00 & 1997 & NTT  & V(5), I(5) & 1 \cr
350-38 (Haro 11)  & 00 34 25.7  & -33 49 49  & 1984 & 2.2m  & B(25), V(15),
Gunn $\it i$(20) & $<$1 \cr
&&& 1988 & 1.5D & Gunn $\it r$(90) & 1 \cr
&&& 1989 & 1.5D & \ha (20) & 1 \cr
400-43  & 20 34 31.0  & -35 39 42   & 1984 & 2.2m & \ha (60)& 1  \cr
&&& 1989 & 2.2m  & B (30), Gunn $\it r$ (90), \ha (90) & $<$1 \cr
&&& 1995 & NTT  &  \ha (30), H$\alpha$-continuum (55) & 1.5 \cr
400-43b  & 20 34 31.0  & -35 39 42  &  1989 & 2.2m  & B (5), Gunn $\it r$ (5) & 
$<$1 \cr
480-12  & 02 52 32.8  & -25 18 49  & 1984 & 2.2m  & B(25), V(20), \ha (20)& 1 
\cr
&&& 1989 & 2.2m  & B (30), V (20), Gunn $\it r$ (45) & $<$1 \cr
\noalign{\smallskip}
\hline
\end{tabular} \\
\end{flushleft}
\end{table*}

\subsection{Target selection}

In this paper we focus on luminous BCGs. Such BCGs are likely to 
have deeper central potential wells and therefore should be more 
capable retaining the gas than the true dwarfs. In
less massive galaxies the metals may be quickly diluted due to
winds from high mass stars and could therefore mimic a truly
young galaxy. The massive BCGs are more suited for
dynamical mass determinations and they are massive enough to
show similarities with normal disk galaxies in formation.

There is no commonly-accepted definition of a blue
compact galaxy. We have simply chosen luminous galaxies
with high central surface brightnesses, blue colours and low
chemical abundances from the list of Bergvall \& Olofsson
(\cite{bergvall1}). For quite some time we have observed a number of
such galaxies and chosen a few for a more detailed study. We
have selected those that have metallicities $<$ 15\% of the solar
values, although none of them as low as IZw18. In view of the
discussions concerning the possible existence of young
galaxies in the local universe we also as a second criterion 
rejected objects that showed regularity in the central regions. 
This was done after inspection of CCD images (mostly unpublished data) 
obtained with the ESO 1.5m telescope of about 15 galaxies from 
the Bergvall and Olofsson list. Here we discuss the properties 
of four of these and two of their companions. We hope that by 
making a more detailed study as presented here and in
accompanying papers we can contribute to improving the classification
methods for this type of BCG. Throughout this paper we will assume a Hubble 
parameter of H$_0$=75 km s$^{-1}$ Mpc$^{-1}$.

\begin{table}
\caption[ ]{{\bf Log of near-IR imagery}}
\begin{flushleft}
\begin{tabular} {llll}
\noalign{\smallskip}
\hline
\noalign{\smallskip}
ESO id. & \multicolumn{3}{l}{Integration time (seconds)} \\
  & J & H & K$^{\prime}$ \\
\hline
\noalign{\smallskip}
338-04 & 1000 & 2000 & 3000 \\
338-04b &1000 & 2000 & 3000 \\
358-38 & 1000 & 3000 & 4000 \\
400-43 & 2000 & 4000 & 5000 \\
400-43b & 1000 & 1000 & 2000 \\
480-12 & 1000 & 1000 & 2000 \\
\noalign{\smallskip}
\hline
\end{tabular} \\
\end{flushleft}
\end{table}

\begin{figure*}
\includegraphics{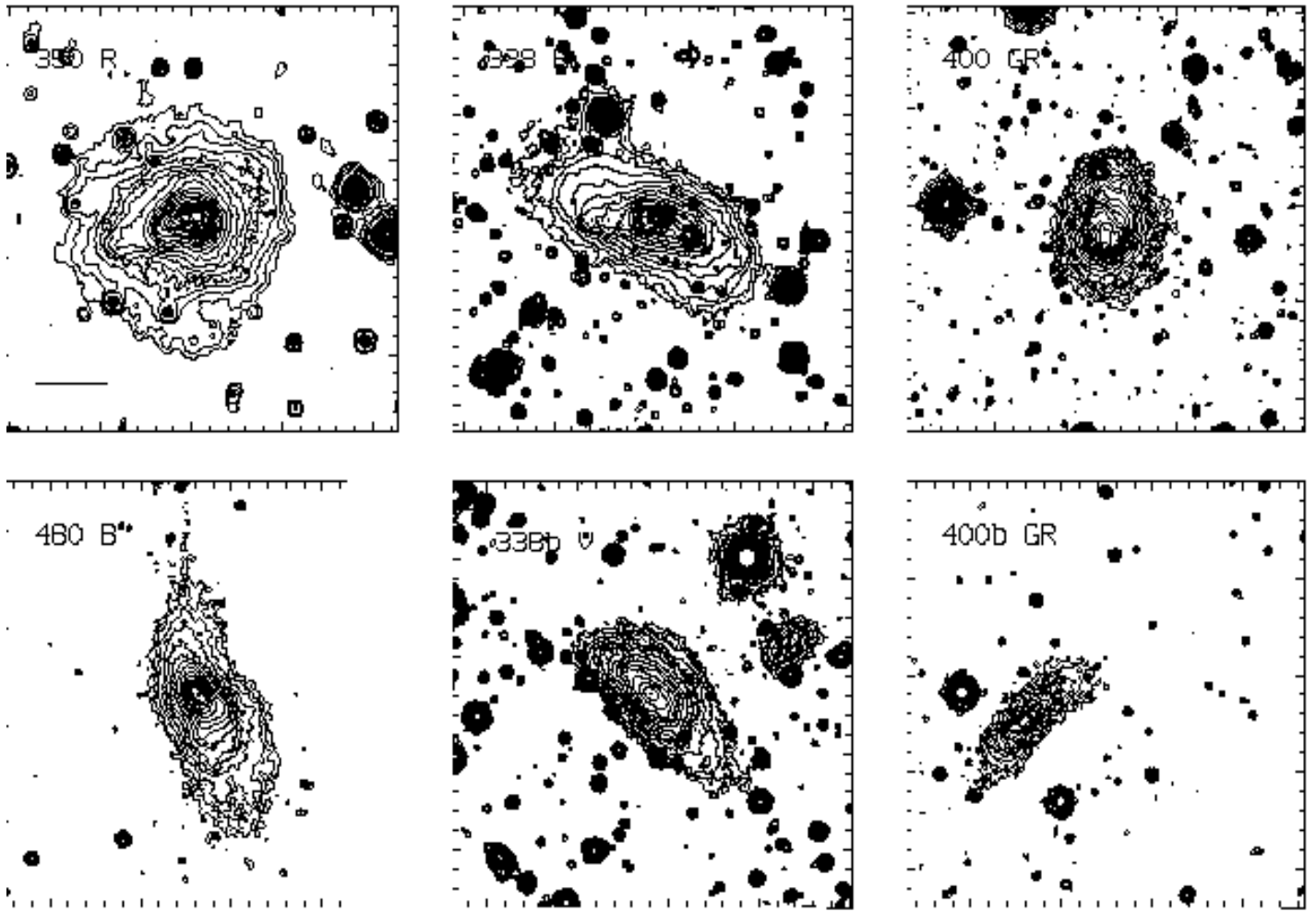}
\caption[]{Optical broadband images of ESO 338-IG04 (Tol 1924-426), 338-IG04b, 
350-IG38 
(Haro 11), 400-G43, 400-G43b and 480-IG12. The filters, indicated after
the abbreviated object name on top of the images, are in the Bessel system 
except those marked GR which is the {\it r} filter in the Thuan-Gunn
system. A median 3x3 pixel filter was applied on the lower flux levels. The
steps between the isophotes are constant on a logarithmic scale arbitrarily 
chosen
for each galaxy such that the details are enhanced.
North is up, east is to the left. The field size is 95"x95"}
\label{broadopt} 
\end{figure*}

\begin{figure*}
\includegraphics{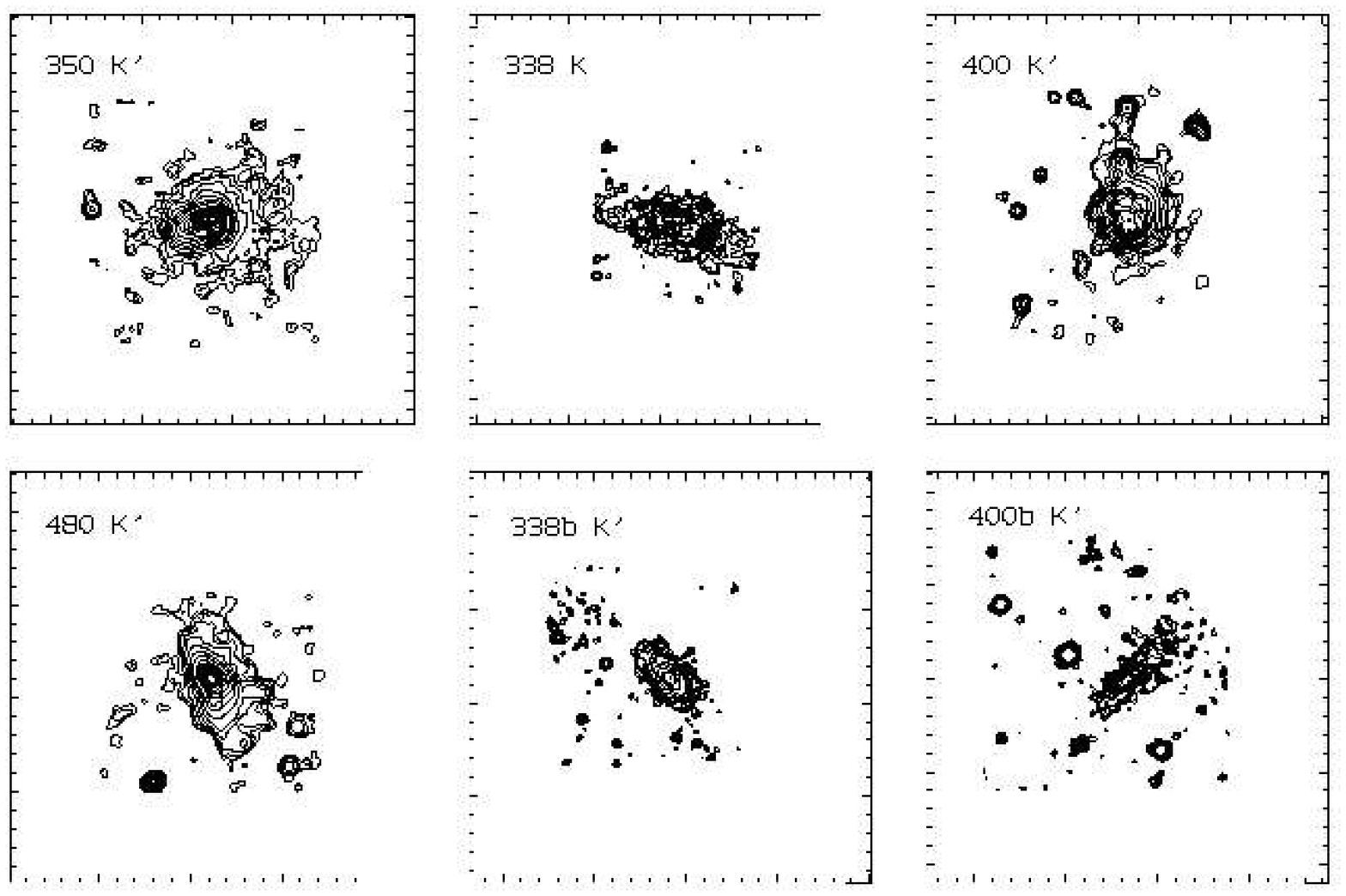}
\caption[]{Johnson $K^\prime$ broadband images of the programme galaxies. The
steps between the isophotes are constant on a logarithmic scale arbitrarily 
chosen
for each galaxy such that the details are enhanced. North is up, 
east is to the left. The field size is 95"x95"}
\label{broadir}
\end{figure*}

 \section{Observations and reductions}

 \subsection{Spectroscopy}

Spectroscopy was carried out at ESO, La Silla, in 1983 and
1984 using the IDS at the ESO 1.5-m and 3.6-m telescope
and in October 8-10 1986 using the EFOSC1 spectrograph/camera
at the 3.6-m telescope, equipped with an RCA CCD
chip. In the first case we used an aperture of 4"x4"
and a spectral resolution of 12 \AA. The spectral
coverage was $\sim$ 3600-7700 \AA. The aperture was centered on the
maximum intensity in the visual of the central region of the
galaxies. The weather conditions were fair. EFOSC1 was
used with the B300 and O150 grisms, with a wavelength
coverage of 3600-7000 and 3600-5590 \AA ~and a dispersion of
230 and 130 \AA mm$^{-1}$ respectively. Here we will only discuss
some of the results obtained with the B300 grism. We used a
slit size of 2 arcseconds. It gives a spectral resolution of $\sim$ 15
\AA. The seeing conditions were poor with a mean seeing of $\sim$ 2
arcseconds. At both occasions standard stars were observed each
night to derive the response curves. La Silla mean extinction
was adopted when correcting to zero airmass. Dome flatfields
were obtained with a Tungsten lamp. 
The spectra were reduced using
the ESO MIDAS software. Final measurements of the spectra
were made using software developed in Uppsala. When
measuring line strengths, the continuum was defined with a
linear approximation and the line with a gaussian
approximation. The H$\alpha$ line was derived from a
decomposition of two Gaussian approximations of the \ha line
and the [NII]$\lambda$6584 line.

\subsection{Optical photometry}

Broad and narrowband optical images were obtained at five
occasions in the period 1983-1997. The broadband filters
UBVRI were in the Cousins system and {\it r} and {\it i} filters also in
the Gunn-Thuan system. Narrowband images were obtained in
\ha and \xhb. The bandwidths of these filters were $\sim$ 70 \AA.
The central wavelengths coincide roughly with the central rest
wavelengths of the lines. Standard stars were obtained during
each night. Colour transformations between Cousins and the Gunn-Thuan systems 
were carried out simply by comparing images of the same objects obtained at 
different occasions in the two different systems and making the zeropoints 
agree with the Cousins system. The colour dependence was thus not taken 
into account. From previous calibrations of colour transformation equations 
(R\"onnback and Bergvall \cite{jari2}) we estimate that the mean errors in the 
colours due to this approximation are less than a few 0.01 mag. The major 
conclusions are based on data unaffected by this. Mean extinction coefficients 
for La Silla were used
in the photometric corrections to zero airmass. The weather
conditions for photometry were average-excellent. The
observations are summarized in Table 1.

\subsection{Near infrared photometry}

JHK$^{\prime}$ observations were carried out with the near-IR camera
IRAC2 at the ESO 2.2-m telescope, equipped with a 256x256 pixel$^2$
NICMOS detector in August 1993. Of the available field
lenses we chose to work with the C lens, which gives a pixel
size of 0.49 arcseconds and a field size of 2 arcminutes. The
weather conditions were good with a seeing of 1-1.5
arcseconds. At least 3 different standard stars each night were
used for calibration. For flatfield corrections we used a superflat
constructed from about 10 object frames where the target galaxy had
been shifted to different positions on the chip between the
integrations. A list of integration times in each filter is found
in Table 2.

\subsection{Luminosity profiles}

Since much of the discussion below focuses on the shape of the
profiles at the faintest surface brightness levels, is it crucial to
have full control of the error estimates that determine the
reliability of the structure parameters. We have good control
of the reductions of the optical images, since the field size is
considerably larger than the galaxies. In the near-IR images the
field size is smaller and the sky background is noisier which
makes the sky subtraction more unreliable. We are
 aware of these problems when we analyze the fainter isophotes
in the near-IR profiles. 

The profiles were derived in the following
way. Firstly we removed hot pixels in the images. This is
automatically done when we stack the images in cases where
we have several exposures. In other cases, where only one
exposure is available, we used the command filter/cosmic in
MIDAS. This procedure efficiently and highly selectively removes hot pixels
and nearby correlated pixels. Secondly, we approximated the sky
background by fitting the sky brightness distribution, measured by
integrating in small boxes placed in regions which appeared empty of
stars, to a 1-3 degree polynomial. We used a degree as low as
possible so that the
residuals looked acceptable. To estimate the uncertainty in the final sky level, 
we calculated the
 median sky value in each box and the standard deviation of the median
 values for all the boxes. We use this as a measure of the error
of the zeropoint of the sky level. This normally dominates the error
sources of the surface luminosities at the faintest isophotes. 
To further check the
stability of the result we in a few cases remade the sky
subtraction. Next step was to remove all stars and
galaxies in the field by hand and flag these regions not to be
included in the following derivation of the profile. Then we
determined the centre, inclination and position angle of the
outermost profiles and integrated the light in elliptical strips
with a width of 1 pixel based on these parameters. We want to point out already
here that the profiles displayed in the diagrams presented below have also
been corrected for inclination assuming an infinitely thin disk.

\section{Spectral evolutionary models }

The spectral evolutionary models (SEMs) we use in the discussions regarding the 
stellar content
come from an in-house model by Zackrisson et al. (\cite{erik}), 
PEGASE2 (Fioc \& Rocca-Volmerange \cite{fioc}) and the model from Worthey
(\cite{worthey}). In the spectral analysis we also use a model by Bergvall
(Bergvall and R\"onnback \cite{bergvall4}). The Zackrisson et al. models are 
based on stellar 
evolutionary tracks with the major contribution from the Geneva group,  
synthetic stellar spectra from the compilation by Lejeune et al.
(\cite{lejeune}) and a nebular component obtained from the Cloudy model
(Ferland \cite{ferland}). In sect. 6 we use these models to compare between 
the  
predicted and observed broadband colours of the halo. We show below that the
contribution from ionised gas to the halo light is fairly small, also in the
R band, and that the uncertainties due to the nebular emission are not
significantly affecting the derived results. This facilitates the analysis and 
improves the reliability of the method when used for
information about the star formation history.

It is important to clarify that we do not intend to
discuss the ionisation structure, sources, metallicities and so on
in any detail here but rather the reverse - we are content with
the situation that we do not need to worry about this
component in the halo. If we had the ambition to discuss the
star formation history of the central burst the situation would
be much more complicated due to the unknown effects of dust,
subcondensations in the gas and Lyman photon leakage, just to
mention a few problems.

\begin{figure*}
\includegraphics[width=18cm]{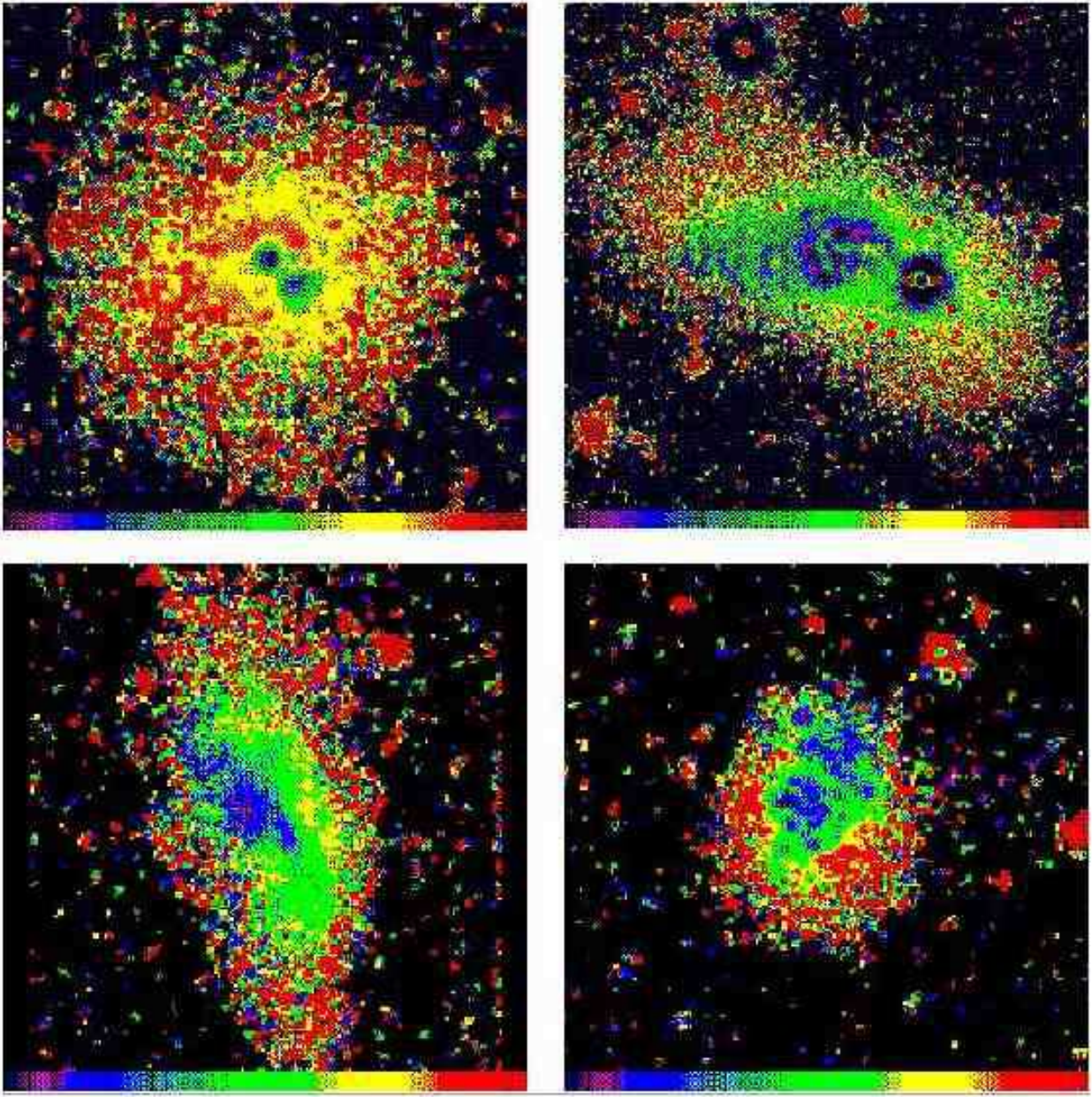}
\caption[]{B-R colour-index maps of ESO 350-IG38 (upper left), 338-IG04 
(upper right), 480-IG12 (lower left) and 400-G43 (lower right). The colour
coding follows true colours with blue representing young star forming regions.
The dynamical range has been chosen to enhance differences: for 350-38
and 338-04 the range is B-R=0.2--1.2, while for 480-12 and 400-43 it is 
B-R=0.2--1.0. 
North is up, east is to the left. The  size  of each image is $1\arcmin 
\times 1\arcmin$. The feature $\sim 15\arcsec$ west of the centre in 
ESO~338-IG04 is a residual from the bright foreground star.} \label{broadcol}  
\end{figure*}

 \begin{figure*}
 \resizebox{\textwidth}{!}{\includegraphics{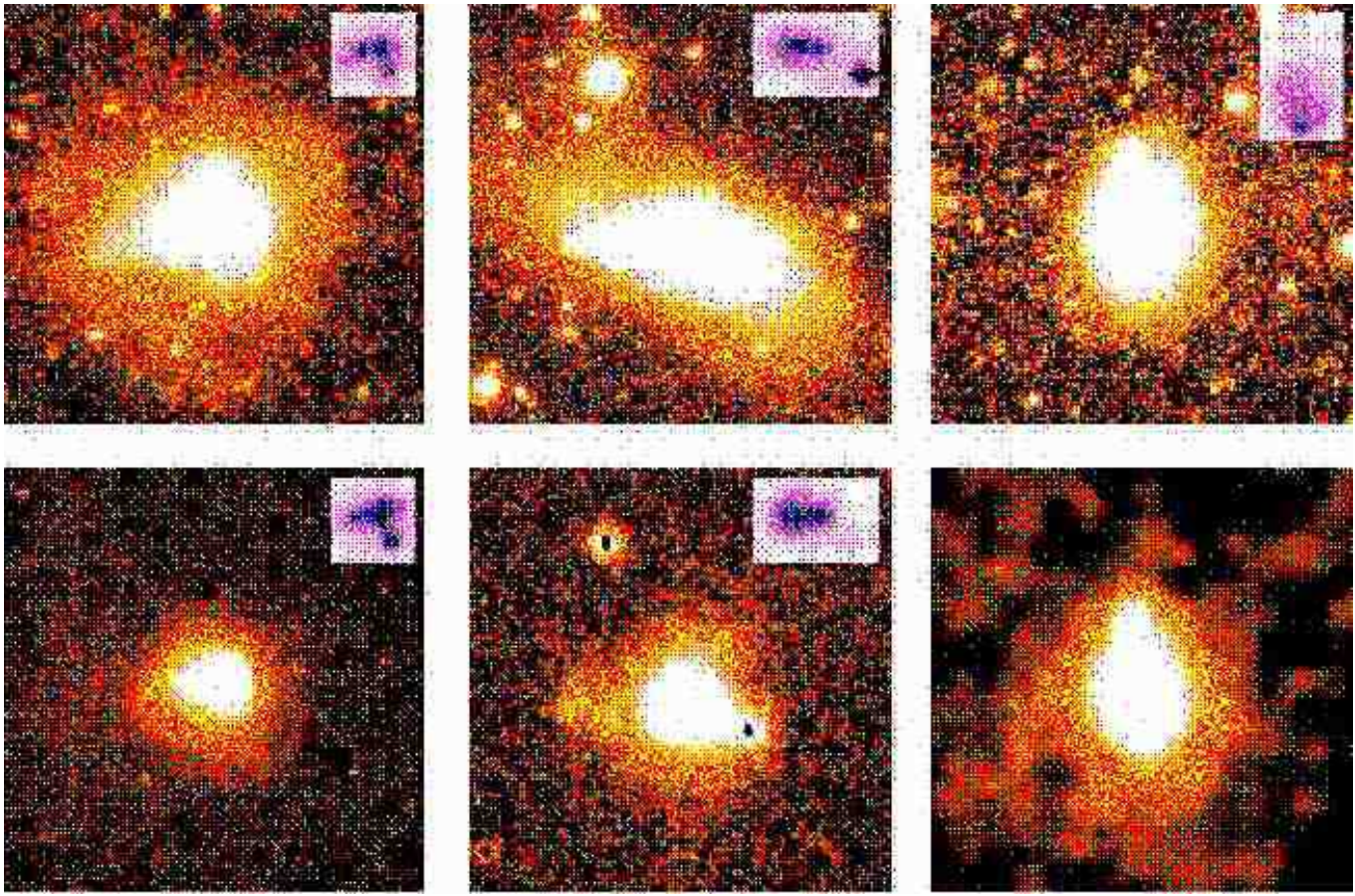}}
  \caption[]{The images shown are from left to right ESO 350-IG38, 338-IG04 
and 400-G43. The upper panel shows deep images in the R window and the
lower panel continuum subtracted \ha images. The inserts in the upper right 
corners 
are displayed
to show the details of the central regions and have the same scale size as the 
larger 
images. The field size is $1.2\arcmin \times 1.2\arcmin$. North is up, east is 
to 
the left. (ESO
1.5m D, MPG/ESO 2.2m and NTT telescopes.)}   
\label{randha}   
\end{figure*}

\begin{table*}   
\caption[ ]{{\bf Integrated surface photometry} The table lists the B magnitude 
and broadband
colours obtained from integrating in elliptic rings of position angle  PA
(degrees), inclination angle $\it i$ (degrees) and a bin width of 0.47
arcseconds  out to a radius, R$_{Ho}$", corresponding to the position when the
median surface  brightness (uncorrected for inclination) in the ring is close
to the Holmberg surface  brightness, i.e. $\mu_B$=26.5 mag. arcsec$^{-2}$. The 
absolute 
magnitudes have been corrected for Galactic extinction according to Burstein and 
Heiles 
(\cite{burstein}). $\Delta \mu$
is the inclination correction of the surface brightness in magnitudes, r$_{Mpc}$ 
is the distance in Mpc.}  
\begin{flushleft}   
\begin{tabular} {llllllllllllll}   
\noalign{\smallskip}  
\hline   
\noalign{\smallskip}   ESO id. & PA$^{\circ}$ & $\it i^{\circ}$ & $\Delta \mu$ &
R$_{Ho}$" & r$_{Mpc}$ & M$_B$ & B  & B-V & V-R & V-I & V-J & J-H  & H-$K^\prime$ 
\\ 
\hline \noalign{\smallskip}
338-04  & 162 & 61 & 0.79 & 42 & 38 & -18.9 & 13.98 & 0.40 & 0.11 & 0.11 & 
(0.85)$^1$ & (0.51)$^1$ & (0.25)$^1$ \\
338-04b$^2$ & 135 & 58 & 0.69 & - & 38 & - & - & -& - & 0.58 & 1.58 & 0.57 & 
0.04 
\\
350-38 & 120 & 35 & 0.22 & 24 & 82 & -20.0 & 14.57 & 0.58 & 0.18 & 0.39 & 1.10 & 
0.63 & 0.60 \\      
400-43 & 163 & 40 & 0.29 & 20 & 77 & -19.6 & 14.89 & 0.62 & 0.00 & - & 0.48 & 
0.63 & 0.23 \\      
400-43b$^3$ & 40 & 70 & 1.16 & 26 & 77 & -17.8 & 16.55 & - & - & - & - & 0.38 & 
0.36 \\    
480-12 & 20 & 64 & 0.90 & 34 & 60 & -18.9 & 14.96 & 0.37 & 0.13 & 0.44 & 1.00 & 
0.70 & 0.14 \\        
\noalign{\smallskip}  
\hline   \end{tabular} \\   
\end{flushleft} 
1) The Holmberg radius could not be reached in the near-IR so the given 
colour is valid for a radius of 30 arcsec.
  
2) m$_V$=15.0, M$_V$=-18.3 at $\Phi$=50"

3) B-R=0.40, B-J=2.09
\end{table*}

\section{Results}

\subsection{Morphologies}

In Fig.~\ref{broadopt}, ~\ref{broadir} and ~\ref{broadcol} we present 
images and colour-index maps of the programme galaxies. The near-IR 
images are strongly limited by the small field size of the detector 
but in fact the profiles reach quite far out from the  starburst
region and it is hard to find corresponding data in the literature. 
In contrast  to normal late type galaxies the BCGs show a strong 
morphological similarity between the optical and the near-IR. Since
the visible light is dominated by young stars, the morphological
similarity suggests that young stars (e.g. red supergiants) dominate   
the near-IR light as well.  At intermediate light levels irregular 
structures, reminiscent of whisps, shells and tails show up.
  At fainter light levels a somewhat more 
regular structure is seen but at the faintest levels again we see 
no firm evidence of an equilibrium system.

\subsection{General characteristics of the programme galaxies.}

Photometry and basic spectroscopic data of most of the galaxies
were presented by Bergvall and Olofsson (1986). A study of 
the velocity fields in \ha of all the programme galaxies 
has been published (\"Ostlin et al. \cite{goran2}, \"Ostlin et al. 
\cite{goran3}).

  \subsubsection{ESO 338-IG04 (Tololo 1924-416)}

\begin{figure}
\includegraphics[width=9cm]{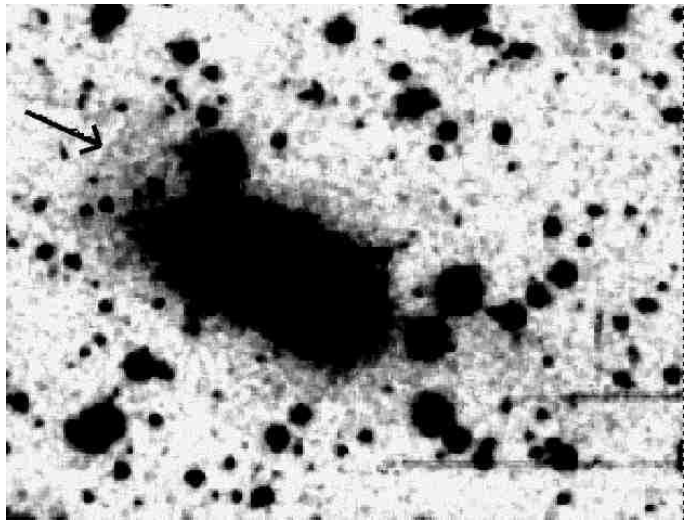}
 \caption[]{A deep, median-filtered image of ESO 338-IG04 in V. The arrow on the 
left side
indicates what might be a remnant of a disk participating in a merger. The field 
size is $2.9\arcmin \times 2.2\arcmin$. North is up, east is to the left. ESO 
NTT.}
 \label{deep338}
  \end{figure}

\begin{figure}
  \includegraphics[width=9cm]{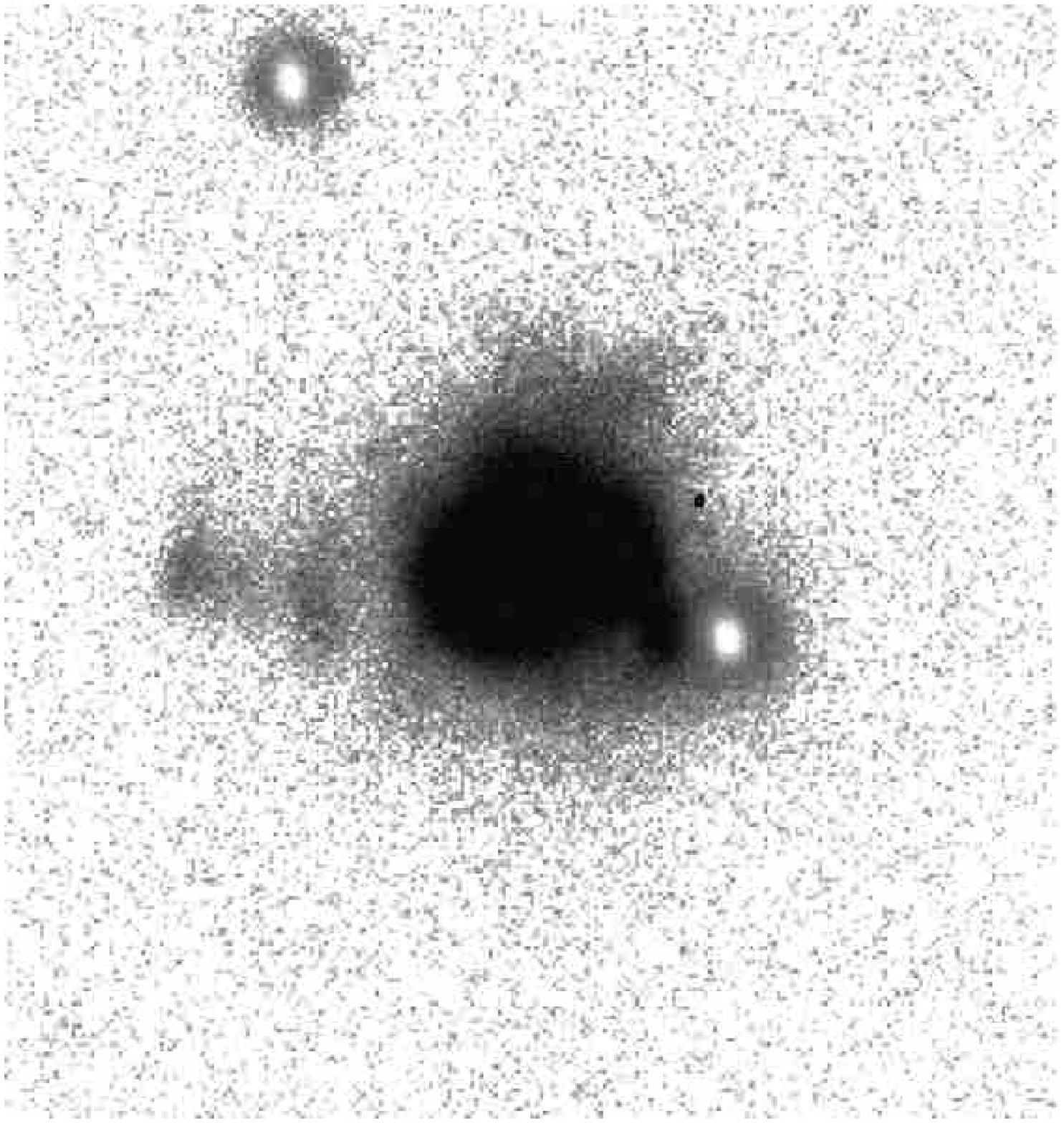}
  \caption[]{A map of the equivalent width of \ha in emission of ESO 338-IG04. 
The continuum subtraction was made using a \ha filter tuned at a different 
redshift.
The displayed range in \wha is $\sim$ 0-500~ \AA~ and the darker parts 
correspond 
to the largest values. The values in 
the halo are $\leq$ 10 \AA. The two blobs with bright centers are residuals
from bright stars. The field size is $1.1\arcmin \times 1.1\arcmin$. ESO NTT.}
  \label{e338wha}
  \end{figure}

This well-known BCG has been discussed by us in two previous papers 
(Bergvall et al. \cite{bergvall0}, \"Ostlin et al. \cite{goran1}). 
 The optical light is dominated by 
the central irregular starburst, clearly resolved into compact star 
clusters in HST images (Meurer et al. \cite{meurer}, \"Ostlin et al. 
\cite{goran1}).  At faint isophotal levels the morphology is that 
of a warped disk. In a filtered image (Fig.~\ref{deep338}), where 
we have made an effort to enhance the faintest structures, a sharp
edge-like  structure is seen in the northeastern part of the disk, 
possibly indicating a warp or a shell. \ha kinematics derived from 
the Fabry-Perot spectroscopy (\"Ostlin et al. \cite{goran3}) reveal 
what may be interpreted as two dynamically separate systems, as if 
the galaxy went through a merger. Fig.~\ref{broadcol} shows a clear 
distinction in colour between the starburst and the surrounding 
host galaxy. Moreover, there is a blue tail, with peculiar kinematics
(\"Ostlin et al. \cite{goran3}) extending eastwards from the starburst.
In the  \ha image (Fig.~\ref{randha}, see also Fig. \ref{e338wha}) 
we see structures extending   perpendicular to the plane, probably 
due to nebular  emission from bipolar outflows.  

This galaxy is of particular interest because it has been shown to
host a system of globular clusters, found to have a range of ages, 
from young ones to old  (\"Ostlin et al. \cite{goran1}). The distribution 
in age indicates that the galaxy has had a few active starburst periods 
in the past. The maximum duration of the current burst is estimated from 
the HST data to be less than 100 Myr or more probable 50 Myr, and the 
last previous major burst probably occurred about 2 Gyr ago. The centre of
the   distribution of the globular clusters is offset with respect to  the
starburst but agrees with that of the underlying red component.  

Bergvall (\cite{bergvall0}) reported an increase in the optical brightness 
between 1979 and 1983 and Gondhalekhar (\cite{gondhalekar}) reported an 
increase in the UV brightness during the same period. Our new data from 
1997 and recent data from Doublier (\cite{doublier}) seem to confirm that 
the galaxy is variable in the optical.  A comparison between our near-IR 
photometry from different epochs shows no variability. Bergvall and Olofsson 
(\cite{bergvall1}) obtained with a diaphragm of 15 arcseconds $J=13.3 \pm 0.1$, 
$H=12.7 \pm 0.1$, $K=12.5 \pm 0.1$ as compared to $13.2 \pm 0.04$, $12.7 
\pm 0.04$ and $12.4 \pm 0.05$ from our integrated surface photometry of 
the IRAC2 images in a circular aperture. Variability in the galaxy will 
seriously affect the interpretation of the photometry of the central region 
discussed below but should not affect the halo data.
  
\subsubsection{ESO 338-IG04b}

This is a spectroscopically confirmed companion of the former
galaxy (Bergvall unpublished; see also Ostlin et al. 1999). 
Morphologically it could be classified as a late type dwarf 
irregular, with moderate star formation activity.  Contrary to 
ESO 338-IG04 it shows regular kinematics, and a dynamical mass
model shows that it contains dark matter (\"Ostlin et al. 
\cite{goran2} and \cite{goran3}). It may have a smaller companion 
$\sim$ 40" westwards (see Fig. 2).

\subsubsection{ESO 350-IG38 (Haro 11)}

\begin{figure}
  \includegraphics[width=9cm]{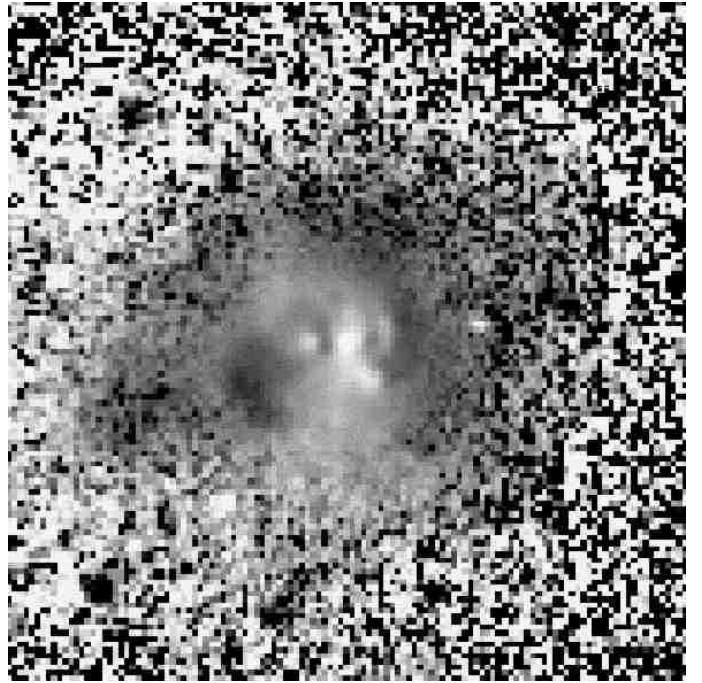}
  \caption[]{A map of the equivalent width of \ha in emission of 
ESO 350-IG38. The range in \wha is 0-700~ \AA~
and the brighter parts correspond to the largest values. The values in 
the halo are about 50 \AA~ except in the southernmost direction where it 
reaches a few hundred. The field size is 1'x1'. ESO 1.5m Danish telescope.}
  \label{e350wha}
  \end{figure}

Three bright condensations are seen in the centre of the galaxy,
which hosts a strong starburst as  evident from the high excitation, 
large emission line equivalent widths and a strong signature of 
emission lines from WR stars in the spectral region around 
HeII $\lambda$4686. Haro 11 is also an extremely hot IRAS source 
(Bergvall et al. \cite{bergvall6}). While the broadband images show 
three condensations in the centre, Fig.~\ref{broadcol} shows 
two blue hotspots. However, the \wha image (fig. \ref{e350wha}) 
shows that the third condensation has the highest \wha value. 
The colourmap shows a gradual transition to redder 
colours with distance from centre. The central bent structure can 
be followed to faint isophotal levels, which together with the
peculiar kinematics, strongly suggest a merger origin of the starburst 
( \"Ostlin et al. \cite{goran3}).

\"Ostlin (\cite{goran4}) revealed the presence of a large number of
globular cluster candidates, similar to the ones in ESO 338-IG04.  
Faint whisps and indications of shell structures are seen in the faint  
outskirts of the mainbody. Approximately perpendicular to the apparent 
major axis (which has a position angle $\sim$ 110$^{\circ}$), the 
\ha distribution is extended, suggesting the presence of  bipolar 
outflows of ionised gas (Fig.~\ref{randha}). Also \wha is enhanced 
south of the centre, in approximately the same direction (Fig.~\ref{e350wha}).

We observed this galaxy in H{\sc i} with the VLA, the Nancay antenna 
and with the Parkes antenna (all unpublished) but only an upper limit 
of the H{\sc i} mass of \ma $_{HI}<$ 10$^8$ \sm could be obtained. With
a \ma $_{HI}$/\ma$_{tot}<$0.01, this galaxy seems to be remarkably 
devoid of neutral hydrogen.
We made an effort to quantify the conditions in Haro 11 on basis 
of observations carried out in the far-IR using the ISO LWS (Bergvall
 et al. \cite{bergvall6}).  We concluded that most of the neutral 
H{\sc i} was located in photodissociated regions. Starting from 
the simplified situation discussed in section 5  we may derive an estimate 
of the mass of the ionised hydrogen gas. From the spectroscopy we calculate 
a central electron 
temperature of T$_e$=13700$\pm$300K and density of n$_e$$\sim$ 10 cm$^{-3}$. 
The \ha luminosity is 3.2 10$^{35}$ W. From this number we derive a
mass of the ionised gas inside 2 scalelengths of 10$\pm$1 10$^8$ 
\sm, assuming a filling factor of 0.01-0.1. The mean HI mass of BCGs of the size 
of 
Haro11 is
8.1$^+_-$0.2 10$^8$ \sm (Gordon and   Gottesman \cite{gordon}). In Bergvall
et al (\cite{bergvall6}) we estimate the fraction of molecular gas and the
mass of the gas in the photodissociated regions to be $\geq$ 3 10$^8$ \sma.
These numbers show that a major fraction of the gas may be in ionised and
molecular form.

  \subsubsection{ESO 400-G43}

\begin{figure}
  \includegraphics[width=8.5cm]{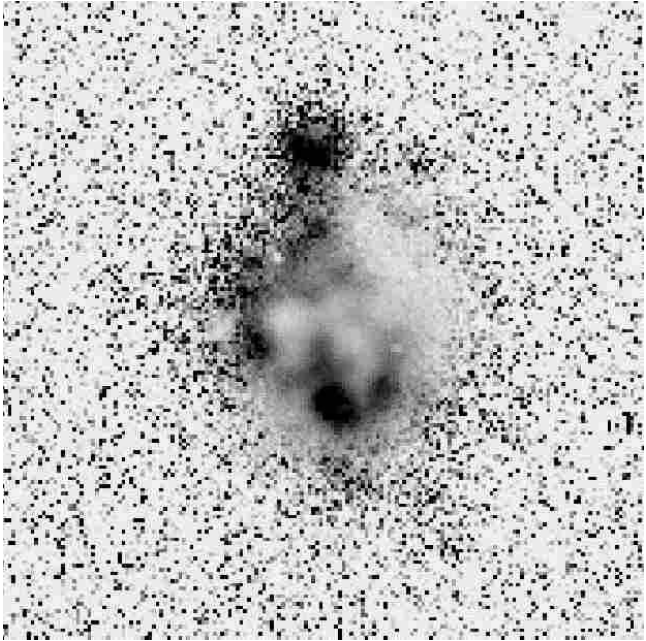}
  \caption[]{A map of the equivalent width of \ha in emission of 
ESO 400-G43. The continuum subtraction was made using a \ha filter tuned at a 
different redshift.
The displayed range in \wha is $\sim$ 0-500~ \AA~ and the darker parts 
correspond 
to the largest values. ESO NTT.}
  \label{e400wha}
  \end{figure}

  As Fig.~\ref{broadcol} shows, this galaxy has a blue irregular, clumpy
central region and a red regular halo. 2 arcminutes east of ESO 400-G43 
is a companion galaxy, here named ESO 400-G43b, detected from the HI 
observations of the main component (Bergvall \&   J\"ors\"ater 1988).  
J\"ors\"ater  and Bergvall (unpublished) observed the main component at the
VLA in the 1415, 4885 and 8414 MHz windows. While the continuum slope in 
the low frequency range agrees with that of Bremsstrahlung radiation
from the ionised gas, a  component  with a steep spectral index   $\alpha 
\sim$ -1.2 (S$^{\nu} \propto \nu^{\alpha}$) shows up near the centre. 
Possible sources are radiosupernovae, SN remnants or possibly a low-mass AGN.
Fig.~\ref{randha} shows a median filtered \ha image of  the galaxy. The 
galaxy has a very extended \ha halo that can   be followed out to 4-5 kpc, or 
$\sim$ 2 scalelengths. The \wha map is shown in Fig. \ref{e400wha}.
A longslit spectrum across the galaxy indicates that 
the extinction is close to zero throughout the central region (Fig.~\ref{hahb}).  

The dynamics and H{\sc i} content of this galaxy were discussed by Bergvall
and J\"ors\"ater (\cite{bergvall2}). One of the purposes with that 
investigation was  to use  dynamical information to set upper limits 
to the mass of older stellar generations. This mass determination is 
particularly important  since it is derived from an HI rotation curve 
reaching far outside the optical extent of the galaxy and as such is 
unique in our sample.  The H{\sc i} mass is $\sim 5\times10^9$ 
\sm and for the companion  $4\times10^9$ \sma. A smaller cloud with a mass
of approximately $1\times 10^9$ \sm and no optical counterpart was detected 
NW of the main component, suggesting that we may be witness to a 
merging of a smaller group of galaxies and gas clouds. From best fits 
to the rotation curve of the optically bright central region it was 
found that the mass was $\sim$ 1.3~10$^9$ \sm, close to the photometric 
mass estimate of young stars 
and ionised gas. Moreover, when corrected for internal extinction 
(using the \xha/\hb ratio) the colours (U-B$\sim -0.8$; B-V$\sim$0.1) 
correspond to a young stellar population. These observations led 
Bergvall and J\"ors\"ater to suggest that  ESO\,400-G43 could be
 a galaxy still in the state of formation. They also 
showed that dark matter  dominates the dynamics at  
$r > 15$ kpc and that the total mass was $> 10^{10}$ \sma, if dynamical 
equilibrium was assumed. A distorsion in the transition between the 
stellar disk and the H{\sc i} was interpreted as  either an effect of 
gas infall or a warp. \"Ostlin et al. (\cite{goran2}, \cite{goran3}) 
mapped the central kinematics in \ha and confirmed the complexity of 
the velocity field in the nort-eastern part of the central disk,
making dynamical mass determinations of the centre more uncertain.

  \begin{figure}
  \includegraphics[width=9cm]{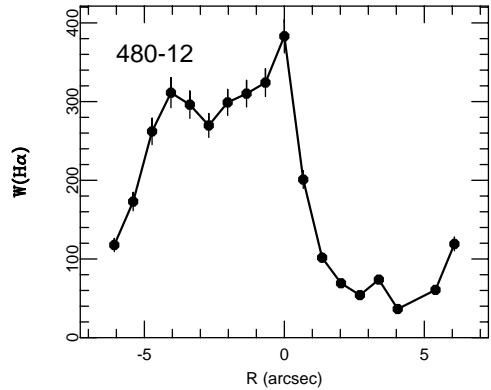}
  \caption[]{\wha as derived from a slitspectrum across the central region of 
ESO 480-IG20 in a position angle of 20$^{\circ}$. The error bars are estimated 
1 $\sigma$ mean errors.}
  \label{e480wha_sp}
  \end{figure}

  \subsubsection{ESO 400-G43b}

  This is the companion of the previously discussed galaxy. It has
  properties of an extensively ionised dI galaxy embedded in an \ion{H}{i}
  cloud with mass $4\times 10^9$ \sma, and a \ml ~ratio that is normal 
  for dIs. Both the spatial extent and equivalenth width of the \ha 
  emission are considerably smaller than for the main component. 
  The excitation, as measured from the [OIII]$\lambda$5507/\hb ratio 
  is moderate, and the galaxy is not a strong IRAS source. Hence, 
  this is not a starburst galaxy.

  \subsubsection{ESO 480-IG12}

  This galaxy has a central condensation and a morphology at the 
  fainter levels that resemble a warped, distorted disk. 
It is the brightest component in a chain of galaxies (see \"Ostlin et al. 
\cite{goran3}, fig. 7). We obtained spectra of three of these (unpublished).
 They all  have high velocities and thus probably do not belong to the system. 
The colour index map (Fig.~\ref{broadcol}) shows a blue spiral-like structure
slightly off centre.   ESO 480-IG12 has the lowest \wha of the BCGs
studied (buth higher than the two companions).   Irregular, warp-like 
structures  are seen in the outer isophotes. Also in this galaxy very faint 
structures stretch out  in a direction perpendicular
  to the disk, indicating that we have bipolar outflows of gas.
  Further support for this comes from our Fabry-Perot spectroscopy 
(\"Ostlin et al. \cite{goran2} 
and \cite{goran3}).

  \subsection{Internal extinction}

  Comparisons between observations of star forming galaxies and
  predictions from SEMs are strongly dependent on the
  reliability of the extinction corrections. In massive star
  forming galaxies this may become a very severe problem since
  we may have a radial dependence on metallicity, age and
  extinction that results in a degeneracy in the information
  obtained, if the wavelength region covered is not sufficiently
  extended. The situation is more favourable in studies of BCG
  as  most observations of BCGs point at a low dust
  content, even in the very central regions (e.g. Terlevich et al. 
  \cite{terlevich}). This is confirmed
  also for the objects discussed in this article as will be shown
  below.

  \begin{figure}
  \includegraphics[width=9cm]{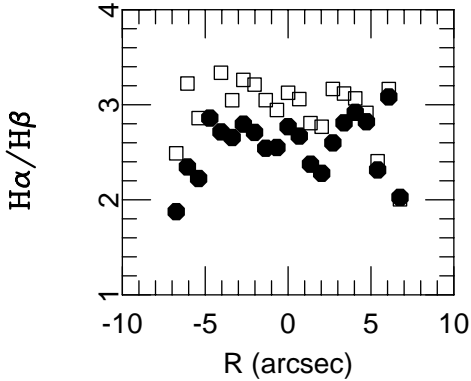}
  \caption[]{The \hahb emission line ratio as derived from a slitspectrum across
  the central region of ESO 400-G43. The open squares are the observed data 
and the filled circles are the data after corrections for underlying
absorption in the Balmer lines.}   \label{hahb}
  \end{figure}

          Since we, at this step, are primarily interested in obtaining
  information about the halo population, the first thing is to
  secure that 1) the extinction is low and 2) the contribution
  from nebular emission can be controlled. We will use slit
  spectra to determine the extinction from the \xha/\hb ratio. Since
  the Balmer emission lines are affected by the Balmer absorption
  lines from the young stellar population, they have to be
  corrected for this before they can be used for extinction
  measurements.
          As a first approximation, the correction was based on \wha in 
comparison with
  predictions from our SEMs. In Fig.~\ref{hahb} we show the results
  from a slitspectrum of one of the programme galaxies, ESO 400-G43. 
We show both the measured \xha/\hb ratio and the same
  ratio after correcting for underlying absorption based on \xwha.
  From the diagram we see that the corrected \xha/\hb ratio
  several times falls significantly below the theoretical ratio
  based on Brocklehursts data (Brocklehurst \cite{brocklehurst})
  which we approximately express

 $$ F(H\alpha)/F(H\beta) = 4.825 - 0.478 \cdot log(T_e) \eqno (4)$$

          What is the reason for this deviation? When comparing
  with slit spectra from other observing runs, covering the same
  region of the galaxy, we can convince ourselves that there are
  no problems with the calibration. The problem is probably due
  to what seems to be an erroneous correction for underlying 
absorption. It seems
  that the correction should be smaller than what we have used.
  The most likely explanation is that we have encountered
  the problem discussed above, i.e. that we underestimate \wha
  either because of leakage along optically thin paths from the
  centre and outwards, or because much of the light from the HII
  region actually falls outside the slit. The second explanation
  seems to be most probable.
          A more realistic correction is to use the colours of the
  starburst to select a best-fit SEM, after iteratively correcting
  for underlying absorption based on the \xha/\hb ratio. From
  this model we then obtain the equivalent widths we need.
  These turn out to be intermediate between the no correction and
  the correction based on \xwha. We will use this method to
  determine the chemical abundances as discussed below. The
  conclusion as regards the extinction in the case we discussed
above is that it is
low (E(B-V) $<$   0.1 mags.) and shows little variation across the galaxy. We
  cannot say for certain that it does not increase in the halo but
  we see no indications of this in the two colour images. We
  will therefore assume that the light from the halo is fairly
  unreddened. That is, when we derive the parameters of the
  luminosity profiles we will make a simple correction for
  internal reddening, assuming that the optical depth is
  proportional to the disk thickness where the zeropoint is taken
  from the derivation of the extinction from the central region.
  The corrections will be quite small however.

  \subsection{Chemical abundances}

  The chemical abundances of nitrogen and oxygen of all galaxies except 
ESO~338-IG04 were derived both from  IDS spectra and EFOSC1 spectra. 
For ESO~338-IG04 Bergvall (\cite{bergvall0}) derived 
$12+\log(\mathrm{O/H})=8.07$,
whereas Masegosa et al. (\cite{pepa}) and Raimann et al. (\cite{raimann})
both got $12+\log(\mathrm{O/H})=7.92$.

 The intensities of 
the most prominent emission lines relative to \hb of the remaining three
BCGs and the two companions are presented in Table 4.  For the
  brighter central regions of the galaxies the calculation of the
  emissivity of the O$^{++}$ zone was based on a determination of the
 electron temperature from the [OIII]$\lambda 4363$/[OIII]$\lambda\lambda
 4959,5007$ emission-line ratio. From the first calculation of T$_e$ we derived
  the theoretical \xha/\hb recombination value (Brocklehurst 
\cite{brocklehurst})
  We then made a reddening correction according to above and
 derived a new T$_e$. In this way the extinction and the
  temperature in the O$^{++}$ region were iteratively determined.
          In order to calculate the total oxygen abundance one would
  like to know also the temperature of the O$^{+}$ zone. Since useful
  lines for such a calculation are either too weak or outside the
  spectral window we calculate this temperature from the
  metallicity dependent relations derived by Vila-Costas and
  Edmunds (\cite{vilacostas}) on basis of Stasi\'nskas (\cite{stasinska}) 
models.
 If T$_e$ is
  expressed in units ot 10$^4$K, the equation we use in this case is

  $$(T_e(O^+))^{-1} = 0.55(T_e(O^{++}))^{-1}+0.36  \eqno (5)$$

\begin{figure*}
  \includegraphics[width=18cm]{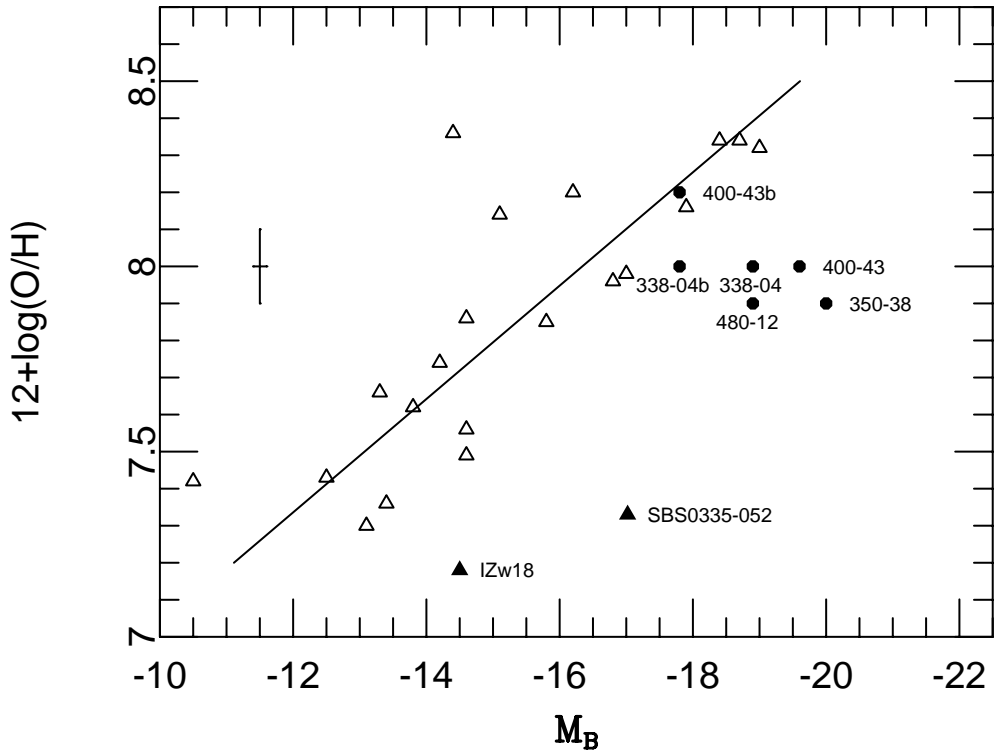}
  \caption[]{The oxygen abundance of the target galaxies as function
  of absolute blue magnitude. The open triangles are local group
dwarf galaxies and the full drawn line is the regression line. Data are from
Skillman (\cite{skillman}). Also shown are the positions of two 
young galaxy candidates, IZw18 and SBS0335-052.}
  \label{mz}
  \end{figure*}

 The total oxygen
  abundance was then calculated from the [OII]$\lambda 3727$ and
  [OIII]$\lambda 4959,5007$ lines using atomic constants from Mendoza
  (\cite{mendoza}). The derived abundances are all $\sim$10\% solar for
  the BCGs and slightly larger for the companions. In Fig. \ref{mz}
we compare the metallicities with the metallicity-luminosity relationship
derived for the local group galaxies (Skillman \cite{skillman}). We
note that the 4 luminous BCGs, but not the companions, deviate 
significantly from the relationship.  The deviation of our
most extreme cases from the Skillman et al. line is the same as for
the two young galaxy candidates, IZw18 and SBS\,0335-052.

\begin{figure}
  \includegraphics[width=9cm]{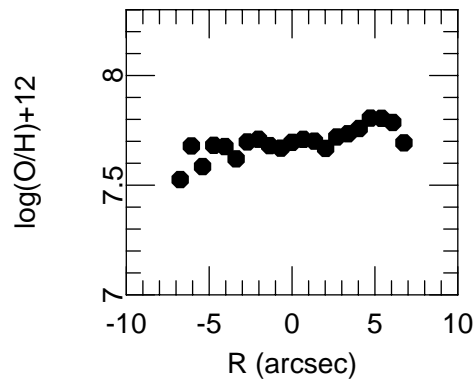}
  \caption[]{The oxygen abundance of ESO 400-G43 as a function
  of position along the disk. The position angle is 110 degrees.
  The abundances are derived from McGaugh's empirical relations
  (McGaugh 1991,1994), utilizing the [OII], [OIII] and \hb lines.}
  \label{oxygen}
  \end{figure}

 A useful piece of information in the interpretation of the difference in 
metallicity
 between the populations in the different regions of the galaxies
is the size of possible abundance gradients in the ionised gas across the
centre and out into the halo. To check the homogeneity of the oxygen
abundances over   the starburst regions we have to work at faint levels in the
  spectra, where the standard method cannot be used. For this   purpose we
applied the empirical relations based on the  
([OII]$\lambda$3727+[OIII]$\lambda \lambda$4959,5007)/\hb line ratio (Pagel et
 al.   \cite{pagel1}, Pagel et al. \cite{pagel2}, Edmunds and Pagel
\cite{edmunds},  Skillman   \cite{skillman}). We are not concerned in this  
article with systematic differences between the abundances   derived from the
temperature sensitive method and the empirical   method since we only wish to
set limits on the amplitude of   the fluctuations in the abundances. As an
example the result   from ESO 400-G43 is shown in Fig.~\ref{oxygen}. We see
that the   variations in the abundances are quite small, and within
the accuracy of the empirical method. The central oxygen
abundances are presented in Table 4 along with the nitrogen abundances as
derived from the [NII]$\lambda$6584 line and the assumption N/H =
(N$^+$/H$^+$)/(O/O$^+$).

  \begin{table*}
  \caption[ ]{{\bf Emission line fluxes and derived properties of the nebular 
gas.} 
The top line gives the ESO numbers of the target galaxies and the second
line the aperture that the fluxes are referred to, centered on the brightest
part of the galaxy in the visual region. Line fluxes are relative to \hb. 
F is the measured emission line flux relative to H$\beta$ and F$_c$ is the 
extinction 
corrected emission line flux relative to the H$\beta$ line flux corrected for 
underlying absorption. n$_e$ is the electron density derived from the [SII] 
lines.
Typical errors in the fluxes are 5\% for the strong lines and 10\% for the 
weaker
ones. Oxygen and nitrogen abundances are in log. units. 
The mean errors in the abundances are typically 0.1-0.2 dex.}   
\begin{flushleft}   
\begin{tabular} {lllllllllll}  
\noalign{\smallskip}   
\hline   
\noalign{\smallskip} Line id. & 338-04b & & 350-38 & & 400-43 & & 400-43b & & 
480-12 & \\
\noalign{\smallskip}  & 2"x2" & & 4"x4" & & 2"x0.7" & & 4"x2" & & 4"x4" & \\  
\noalign{\smallskip}  & F & F$_c$ & F & F$_c$ & F & F$_c$ & F & F$_c$ & F & 
F$_c$ 
\\  
\noalign{\smallskip}   
\hline
\noalign{\smallskip}
\lbrack OII\rbrack $\lambda$3727    & 1.73 & 1.74 & 1.76 &  2.02 & 4.68 & 4.39 & 
4.82 & 3.98 & 2.40 & 2.54\\ 
\lbrack NeIII\rbrack $\lambda$3869  & 0.30 & 0.30 & 0.18 & 0.20 & -  & - & - & - 
& 0.36 & 0.37 \\   
\lbrack OIII\rbrack $\lambda$4363   & 0.033 & 0.030 &  0.050 & 0.051 & 0.058 & 
0.054 & - & - & 0.067 & 0.064\\
HeI $\lambda$4471                    & 0.059 & 0.053 &  0.039 & 0.039 & -  & - & 
- & - & 0.043 & 0.041 \\
\lbrack OIII\rbrack $\lambda$4959   & 1.25 & 1.02 & 1.19 & 1.05 & 1.14 & 1.05 & 
1.00 & 0.83 & 1.46 & 1.22\\
\lbrack OIII\rbrack $\lambda$5007   & 3.69 & 3.00 & 3.80 & 3.30 & 3.32 & 3.06 & 
3.24 & 2.68 & 4.64 & 3.86 \\
HeI $\lambda$5875                    & 0.16 & 0.11 & 0.15 & 0.11   & 0.17 & 0.15 
& - & - & 0.14 & 0.10\\
\lbrack OI\rbrack $\lambda$6300     & 0.16 & 0.11 & 0.068 & 0.047 & 0.20 & 0.18 
& 
- & - & 0.073 & 0.050\\
\ha                                 & 4.13 & 2.76 & 4.08 & 2.75 & 3.12  & 2.82 & 
3.42 & 2.82 & 4.12 & 2.75 \\
\lbrack NII\rbrack $\lambda$6584    & 0.13 & 0.08 & 0.78 & 0.52 & 0.13 & 0.12 & 
0.17 & 0.14 & 0.33 & 0.22 \\
HeI $\lambda$6678                    & - & - &  0.0268 & 0.018 & -  & - & - & - 
& 
- & -\\
\lbrack SII\rbrack $\lambda$6716    & 0.54 & 0.36 & 0.25 & 0.16 & 0.39 & 0.36 & 
- 
& - & 0.26 & 0.17 \\
\lbrack SII\rbrack $\lambda$6730    & 0.44 & 0.30 & 0.13 & 0.086 & 0.25 & 0.22 & 
- & - & 0.19 & 0.12 \\ 
\noalign{\bigskip}
T$_{e,\lbrack OIII\rbrack}$ & & 11600 & & 13700 & & 14400 & & - & & 14100 \\
log n$_{e,\lbrack OIII\rbrack}$ & & 2.3 & & 0.8 & & 0.8 & & - & & 1.8 \\
T$_{e,\lbrack OII\rbrack}$  & & 12000 & & 13100 & & 13400 & & - & & 13200 \\
log n$_{e,\lbrack OII\rbrack}$ & & 2.3 & & -0.2 & & -0.2 & & - & & 1.8 \\
$\log(O/H)+12$ & & 8.0 & & 7.9 & & 8.0 & & 8.2$^1$ & & 7.9 \\
$\log(N/O)$ & & -1.5 & & -0.7 & & -1.7 & & -1.7 & & -1.2 \\
\noalign{\smallskip} 
W$_{H\beta}$ (\AA) & 32 & 38 & 65 & 72 & 32 & 38 & 15 & 18 & 43 & 51 \\

F$_{H\beta}$ (10$^{-18}$Wm$^{-2}$) & 5.3 & 12.6 & 416 & 1120 & 39 & 46 
& 4.5 & 5.4 & 167 & 366 \\ 
\noalign{\smallskip} 
\hline  
\end{tabular} \\   
\end{flushleft}
1) Based on the empirical oxygen abundance relationship by McGaugh 
(\cite{mcgaugh1}). 
\end{table*}

  \subsection{Luminosity and colour profiles}

\begin{figure*}
  \includegraphics[width=18cm]{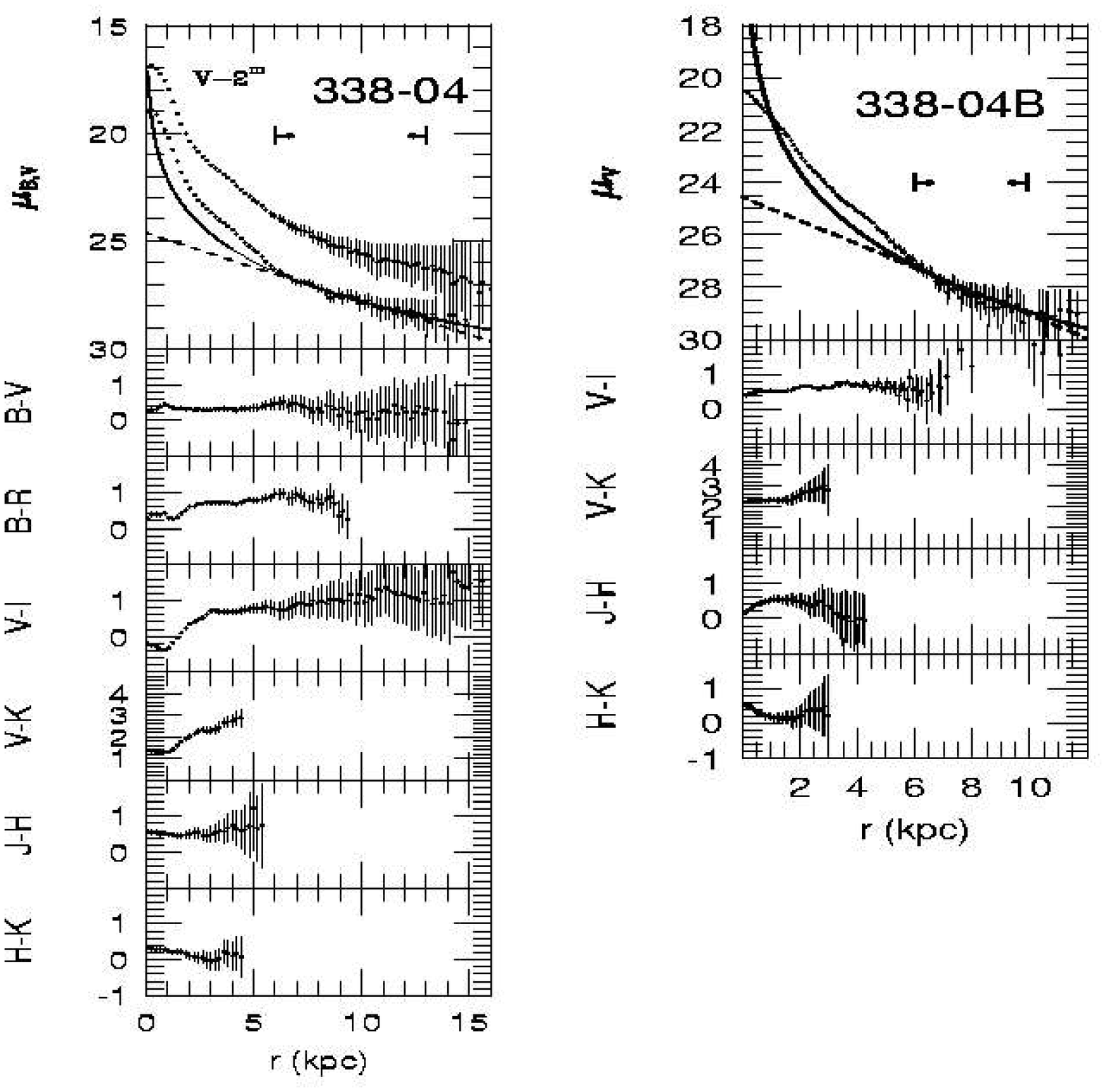}
  \caption[]{The surface luminosity profiles of ESO 338-IG04 and ESO 338-IG04b 
in 
Cousins B and in Cousins/Johnson broadband colour indices.
 The luminosity profiles are shown with two different fits. The dashed line 
shows  
a least square fit to an exponential profile and the solid line a fit to a
Sersic law. Inclination corrections based on a fit to the outer isophotes have
been applied. Likewise reddening corrections based on spectroscopy of the  
central region and assuming the column density of the dust to be proportional
to the luminosity density in the broadband image used to derive the upper
luminosity profile, have also been applied. The error bars are 1$\sigma$
deviations including the most pessimistic estimate of the error in the
zero-point correction of the sky background.}    
\label{338ab}   
\end{figure*}

\begin{figure*}
  \includegraphics[width=18cm]{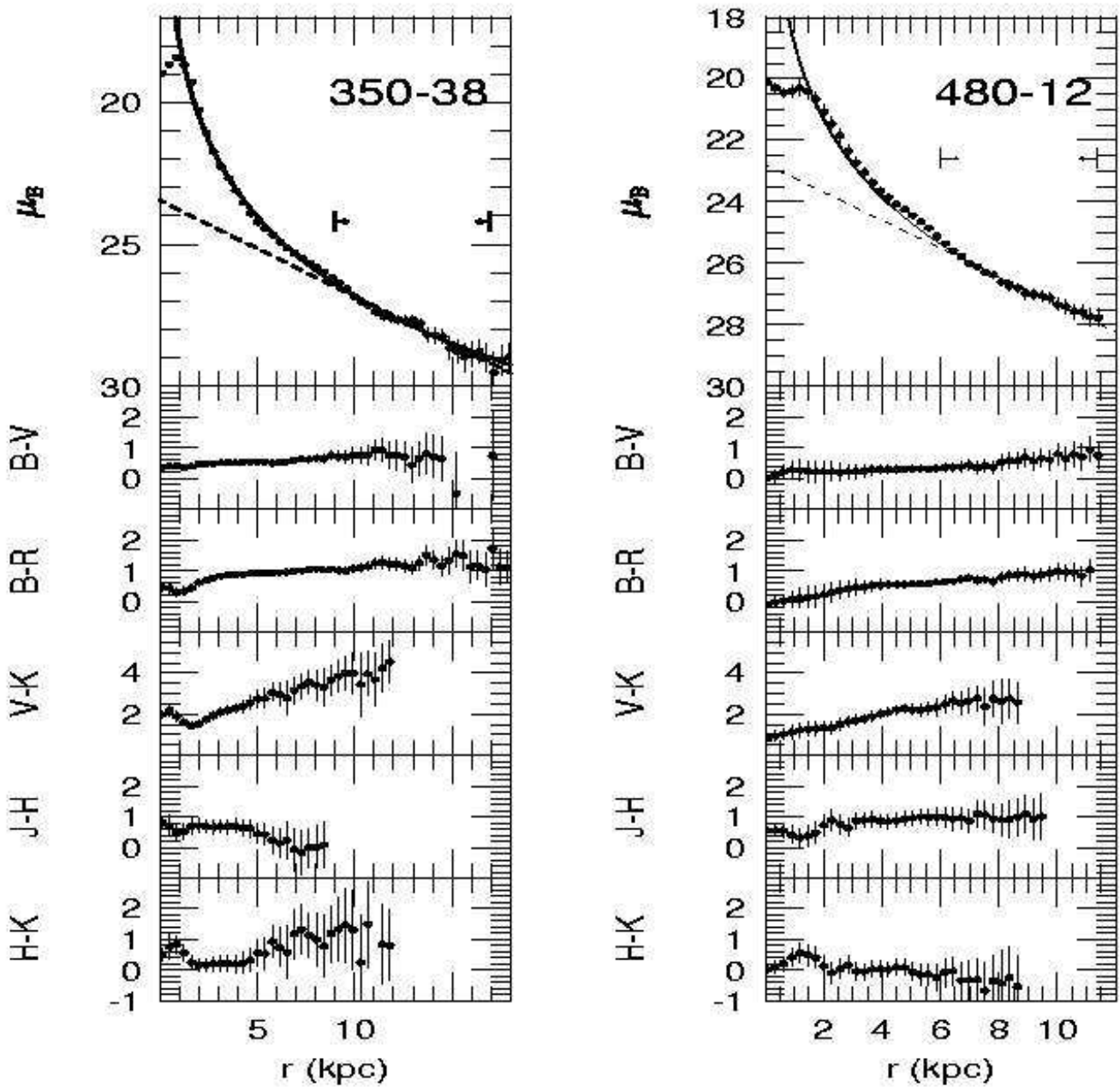}
  \caption[]{Luminosity and colour index profiles of ESO 350-IG38 (Haro 11) and 
ESO 480-IG12.}
  \label{350_480}
  \end{figure*}

\begin{figure*}
  \includegraphics[width=18cm]{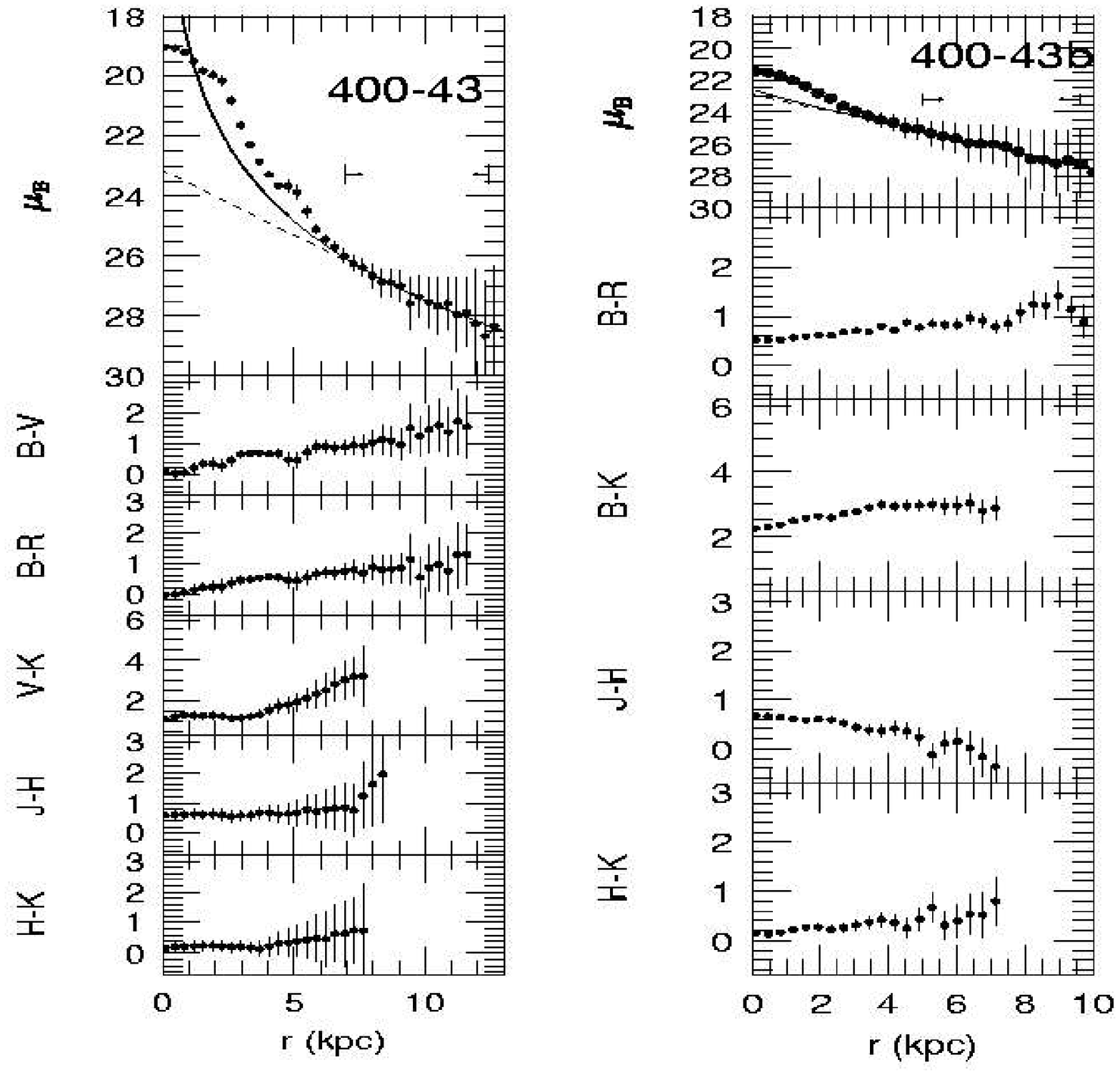}
  \caption[]{Luminosity and colour index profiles of ESO 400-G43 and ESO 
400-G43b.}
  \label{400ab}
  \end{figure*}

  Figures \ref{338ab} to \ref{400ab} show the optical/near-IR luminosity and 
colour
  profiles of the galaxies. We wish to point out that the on-the-spot colours 
within the central 1-2 arcseconds are not reliable due to seeing effects. To
simplify the presentation we will only show the B luminosity profile (except
for ESO 338-IG04 where we also show the V profile reaching a photometric level
of unique depth) and a few colour index
profiles. This will be sufficient for our main goal here which is to  
separate the halo population from the young  burst. Of
course there may be many different components in the galaxy if   several
mergers have occurred but we will show below that it really is possible to use
a two component distinction.  Many investigations of the stellar population in
BCGs are based on BVR surface photometry. A problem when using only BVR data
is that the colour indices, in particular $B-V$, are fairly insensitive to age
if the metallicity is low. It is   thus difficult to use these colours to
discriminate between the burst population and an old halo population. Moreover
the old generation has difficulties to compete with the luminous young
generation. Therefore a combination of optical and near-IR colours is much  
more powerful, as is evident the diagrams. In particular in the   $V-K^\prime$ 
profile
which is shown in the diagram, we see a strong trend towards   redder colours
as we move outwards. Although this trend is present also in the $B-V$ colours it
is much weaker, despite that we have corrected the colours we show in the
diagrams for internal extinction. This is due to the compensating effect
nebular emission has on 'age reddening'. We will call the red stellar  
population the 'halo population', clearly distinguished from the   starburst
population in the centre.

          What we now will do is make a fit to this part of the
  luminosity profiles of the deepest images to see what
  restrictions we can set to the shape of the profile. In general
  (e.g. Sersic \cite{sersic}, Graham et al. \cite{graham}), the luminosity
 profiles
  of galaxies may be expressed as

  $$ I(r)=I_e~exp\big({-b[({r\over r_e})^{1/n}-1]}\big)  \eqno (6)$$

          where r$_e$ is the effective radius, enclosing half of the light
  of the galaxy and b and n are constants. This
  relationship may also be expressed as

  $$ \mu(r) = \mu_0 + {{2.5b}\over{\ln(10)}}\Bigl(r/r_e\Bigr)^{1/n} \eqno (7)$$

          where $\mu_0$ is the central surface brightness. The constant b
  is a function of n and can be expressed as $b_n\approx 2n-0.327$
 (Capaccioli \cite{massimo}).
 n = 4 for a de
  Vaucouleurs profile and n=l for an exponential profile. 
The
  first shape is characteristic of normal ellipticals and the
  second one is typical of disks and dEs. Luminous cluster ellipticals have 
n$\leq$15. There seems to be a continuous
  transition between the two extremes that occurs when going
  from high luminosity Es to low luminosity dEs (Graham et al. \cite{graham}). 

  \subsection{Structural parameters}

  In an effort to disentangle the photometric parameters of host
  galaxies of BCGs, Papaderos et al. (\cite{papaderos}) used luminosity
  profiles obtained from optical broadband images. When fitting
  an exponential disk to the outer part of the profile they found
  that the central surface brightness of this hypothetical disk was
  brighter than that of typical LSBGs ($\mu_{B,0} \geq 23$). The conclusion
  was that LSBGs were rather improbable as precursors of BCGs.
  In the present investigation we reach fainter limiting isophotes
  than they obtained in their work. Our luminosity profiles show that this may
  make a big difference. We see that the scalelength is more or
  less continuously increasing when going to lower surface
  luminosity levels. The faintest luminosity profiles of our
  galaxies reach $\sim$ 29-30 mag arcsec$^{-2}$ (n.b.: after correction
  for inclination). As shown in the figures,
  we get completely different results if we make a fit out to this
  isophote than if we stop at the more commonly used limit,
 $\mu_B\sim$ 25-26 mag arcsec$^{-2}$. In fact we have no problem to reach a
  central surface brightness as low as that of LSBGs by fitting an
  exponential profile to the outer isophotes. The question is however,
  whether a disk really is the best and most appropriate fit.

          To proceed we will make two fits to the outer isophotes,
  one where we leave the parameter n (eq. 7) free and one where we fix it to
n=1 (exponential disks),   in order to facilitate comparisons with LSBGs. The 
range of the fit is set by eye inspection where the halo colours start to become 
significant. In
table 5 we summarize the parameters we derive from fits to the outer
  parts of the luminosity profiles, within the radial ranges as
  indicated as in the figures. The parameters we specify are
  the same as in equation 7. What is striking is that the n
  parameter is very large ($>>$4) when we use good data at large
  distances from the centre. Although the baseline of the fit is short, which 
will make the solution somewhat unstable, we take this as an
  indication that the host galaxy of the burst is of early type since all
galaxies go in the same direction. The fit will not be much worse if we fix n 
to be equal to 4 but we see no reason why we should prefer this value to the
best solution value. We obtain an even better fit (measured by the correlation 
coefficient) to a power-law if we include more data points closer to the centre. 
In the final column of the table we give the maximum
luminosity the power law component can have, simply by assuming that the
brightness follows the power law down to a distance where the model surface
brightness is equal to the observed surface brightness. At shorter radial
distances we have assumed it to be equal to this value. 

There is an exciting consequence of the high n values we derive, should these 
values turn out to be firm, once deeper photometry becomes available. Caon et 
al. (\cite{caon}) found a correlation between n and galaxy luminosity. The 
higher the n value the brighter (and more massive) the galaxy. Similarly, 
Graham et al. (\cite{graham}) found that bright cluster ellipticals typically 
have high $n$ values. The highest best fit n values for the luminous galaxies is 
typically $n\sim 15$. If a high $n$ signals a high mass it could indicate that 
the BCGs in our sample have a high amount of dark matter.

\begin{table*}
 \caption[ ]{{\bf Parameters of the best fits to the luminosity profiles of the 
deepest images}. Range is the range of the disk within which the fit was
made, h is the scalelength of the exponential fit, $\mu_{0,disk}$ is the
central disk surface brightness in mag. arcsec.$^{-2}$,  $\mu_{0,disk}$ is
the absolute B (V for 338-04b) magnitude of the disk, n is the parameter in the
Sersic approximation where n=4 is the de Vaucouleurs law (see text) 
and n=1 is an exponential disk, r$_{e}$ is the effective radius, $\rho$ 
is the correlation coefficient of the fit, $\sigma$ is the mean error in
the correlation coefficient. All these parameters refer to the solid line in 
figures 16-18. M$_{pl}$ is the {\it limiting} absolute 
B (V for 338-04b) magnitude inside the Holmberg radius of the power law 
solution, 
truncated at small radii to the observed surface brightness level. The data are 
based on luminosity profiles that were corrected for internal extinction as 
described 
in the text.}  
\begin{flushleft}   
\begin{tabular} {llllllllllllll}   
\noalign{\smallskip}  
\hline   
\noalign{\smallskip} Galaxy & Filter & Range & h$_{disk}$
& $\mu_{0,disk}$ & M$_{disk}$ & $\rho$ & $\sigma$ & n & r$_{e}$ & $\mu_0$ & 
$\rho$ &
$\sigma$ & M$_{pl}$ \\ 
& & (kpc) & (kpc) & & & & & & (kpc) \\
\hline 
\noalign{\smallskip} 
338-04   & B &  6-11 & 3.3 & 24.6 & -16.6 & 0.968 & 0.008 &  9.8 & 0.921  &  
1.09 
 & 0.973 & 0.005 & -17.7 \\
338-04b  & V &  6- 9 & 2.4 & 24.6 & -15.9 & 0.54 & 0.12 & 19.8 & 8.8e-4 & -39.39 
& 0.55 & 0.11 & -17.5 \\ 
350-38   & B &  8-16 & 3.2 & 23.5 & -17.6 & 0.978 & 0.009 & 19.8 & 3.2e-5 & 
-53.95 & 0.982 & 0.008 & -20.0 \\ 
400-43   & B &  6-12 & 2.5 & 23.1 & -17.5 & 0.981 & 0.010 & 17.1 & 4.2e-4 & 
-38.85 & 0.985 & 0.008 & -19.5 \\ 
400-43b  & B &  5- 9 & 2.3 & 22.9 & -17.5 & 0.978 & 0.014 &  1.1 & 3.76 &  22.54 
& 0.978 & 0.014 & -17.4 \\
480-12   & B &  6-11 & 2.4 & 22.8 & -17.7 & 0.988 & 0.006 & 19.7 & 9.0e-5 & 
-49.18 & 0.996 & 0.002 & -19.3 \\  
\noalign{\smallskip}   \hline  
\end{tabular} \\   \end{flushleft}
 \end{table*}

\section{Nebular emission from the halo region of BCCs}

  \begin{figure}
  \includegraphics[width=9cm]{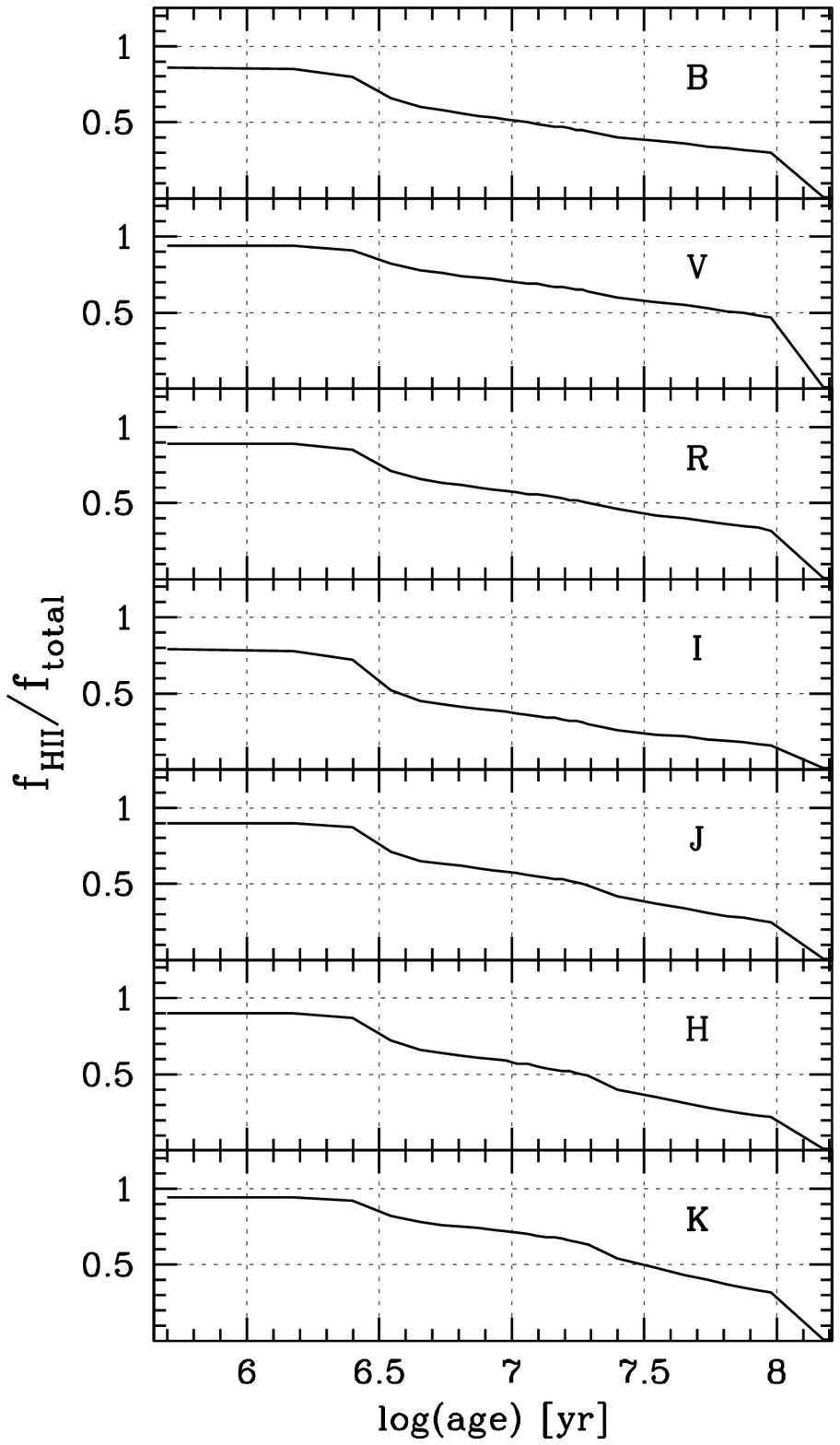}
  \caption[]{The relative contribution from nebular emission to the total 
emission 
(stars+gas) in different broadband windows. The colours are based on a model
of a continuous star formation rate with a Salpeter mass function and a 
metallicity
of 5\% solar.}   \label{hiicol}   \end{figure}

  In a few cases optical colours of the halos of BCGs have been derived that 
have been taken as evidence of an old stellar population (Hunter and
Gallagher \cite{hunter}, Kunth et al. \cite{kunth}). However, one has to be 
aware 
of 
the ambiguities in the
interpretation of these data. The analysis of our programme galaxies shows
that nebular emission may contribute with a considerable amount to the light
from the halo. Possible in situ
ionising sources are young stars or old hot stars, i.e. blue horizontal branch 
stars 
(BHB) or post-AGB stars (PAGB).
If so, it would cause no problem in the following comparisons with SEMs 
predictions, 
since the ionising flux from these stars is taken into account and the
nebular component is included in the models. But, as we will show below, it is
quite possible that gas in the halo is ionised by the central starburst. It is
also possible that the expelled gas from the central star forming regions is
energetic enough to contribute significantly to the ionisation either
mechanically or by conduction.  Considering the current debate about what is
driving the reionisation of the post-recombination universe it is quite
important to identify what is the major  important ionising source in luminous
BCGs to see if they possibly can contribute significantly to this process. 

It seems unlikely that this gas is ionised by stars in situ 
because we see no support of the presence of young star clusters in the
region and the large scale morphology in \ha differs from
that of the broadband colours (Fig.~\ref{randha}). 
We have argued that shocks are important ionisation sources, at least 
close to the central region. It is hard to say how much outflows may
contribute to the ionisation at larger 
distances but the heat input is probably of importance (e.g. Heckman et al. 
\cite{heckman}).

Nevertheless we will now also ask whether it is possible for the young 
starburst to ionise a halo which has an extent several times larger than the
starburst. We made a calculation to the first approximation of how
realistic such a situation would be assuming that the gas is in 
pressure equilibrium and approximately spherically distributed. 
The density distribution was assumed to follow King's
  approximation

  $$n(r) = n(0)\big(1+({r\over a})^2\big)^{-{3\beta}\over 2} \eqno (3)$$

where $a$ is the core radius and $\beta$ is a constant $\approx$ 1. As core
radius we used the typical value of r$_{e}$ derived from the profile fits,
i.e. r$_{e} \sim$ 1 kpc. From the SEMs, based on a Salpeter mass function with a 
mass range of
0.1 to 120 \sm and an  age of 10$^7$ years, we obtain a Lyman photon production 
rate of 
log(N$_{Lyc}$) $\approx$ 54 for a young metal-poor (10\% solar) starburst with a 
luminosity 
typical of our target galaxies. The energy
balance requirement then allows the ionised gas to reach  a radius of 100 kpc 
within
the normal range (e.g. Copetti et al. \cite{copetti}, Martin 
\cite{martincrystal}) of the 
filling factor and central density (f=0.01-0.1 and n$_{e,0}$=10-100 cm$^{-3}$). 
Thus the 
starburst would be capable of ionising the halo out the region we are exploring. 

From our observations we can set limits on the amount of possible excess 
radiation from nebular emission in the halo and how
much it contributes to the light. We have used \ha as a probe to make
reasonable estimates. For ESO 338-IG04 
we constructed a \wha map using as continuum a 2000s deep exposure in an \ha 
narrowband (60\AA ) filter 
designed for a different redshift. The result, Fig. \ref{e338wha}), was
first discussed in Sect. 4.2.1. Here \wha drops below 
10\AA in the centre. For ESO 350-IG38, we used the deep R  image as 'off-line' 
to reach a sufficiently faint surface brightness level. Fig.~\ref{e350wha} 
(also discussed in Sect. 4.2.3) shows 
the resulting \wha map. As is seen, \wha stays at a low level in the
halo. Similarily, as discussed before in Sect. 4.2.4, fig. \ref{e400wha} 
shows \wha of ESO 400-G43. We notice rather high values on 
the northeastern side. For ESO480-IG12 we do not have access to a deep \ha 
image so here we use a slit spectrum. Fig. \ref{e480wha_sp} shows \wha along 
a slit in the major axis position.  We conclude that also in this case \wha 
stays at a  moderate level, around 50-100 \AA ~although there may be 
tendencies for an increase at larger distances. 

We have modelled the colours and fluxes of a pure HII
region in the different wavelengths bands to obtain estimates of how much the 
nebular 
emission will contribute to the different wavelength bands as the starburst
evolves. Fig. \ref{hiicol} shows the relative contribution of the nebular
emission to the total emission from stars and ionised gas. Since \ha dominates
the emission in the R window we can compare the nebular contribution in the R
window with the other wavelength bands and then scale with \ha to obtain the
actual contributions. The effective band width of the Cousins R window is
$\sim$ 1200 \AA. Thus, following the results from the spectroscopy and the
narrowband images, which 
indicated \wha $\leq$ 100 \AA ~, we find that the contribution from nebular 
emission to the total flux in any of the
BRIJHK windows is $<$ 10\% and $<$ 15\% for V. This will have 
a small but not insignificant effect on the
broadband colours when we compare the observations with the predictions. Below
we will return to a discussion about to what extent these data can be used to
constrain the total emission from the outer parts of the halo.

\begin{table}
 \caption[ ]{{\bf Model parameters of the fit to the spectral evolutionary model 
of 
Zackrisson et al. (\cite{erik})} SFH, the star formation history, is assumed to 
be of two types: 
constant (const) or exponentially declining (expo). Z is the metallicity. 
$\tau$ is either the duration 
of the burst or the e-folding timescale of the SFR. n$_e$ is the electron 
density 
of the nebular component, f is the filling factor of the ionised gas and c is 
the 
covering factor of the ionised gas, i.e. the relative amount of nebular emission 
to 
the stellar light responsible for the ionisation.}  
\begin{flushleft}   
\begin{tabular} {llllllll}   
\noalign{\smallskip}  
\hline   
\noalign{\smallskip}
Model  & Model   & SFH & Z & $\tau$ & n$_e$ & f & c \\
set no. & no.& & & (Myr) & cm$^{-3}$ \\
\hline 
\noalign{\smallskip}
1 & 1 & const & 0.001 & 14000 & 100 & 1 & 1 \\
  & 2 & const & 0.001 & 10 & 100 & 1 & 1 \\
  & 3 & const & 0.001 & 100 & 100 & 1 & 0.5 \\
  & 4 & const & 0.001 & 100 & 100 & 0.1 & 1 \\
  & 5 & const & 0.001 & 100 & 1 & 0.1 & 0.5 \\
  & 6 & const & 0.001 & 100 & 1 & 1 & 1 \\
  & 7 & const & 0.001 & 100 & 100 & 1 & 1 \\
  & 8 & expo & 0.001 & 14000 & 100 & 1 & 1 \\
  & 9 & expo & 0.001 & 1000 & 100 & 1 & 1 \\
\smallskip
2 & 1 & const & 0.001 & 1400 & 100 & 1 & 1 \\
  & 2 & const & 0.001 & 10 & 100 & 1 & 1 \\
  & 3 &const & 0.001 & 100 & 100 & 1 & 0.5 \\
  & 4 & const & 0.001 & 100 & 100 & 0.1 & 1 \\
  & 5 & const & 0.001 & 100 & 1 & 0.1 & 0.5 \\
  & 6 & const & 0.001 & 100 & 1 & 1 & 1 \\
  & 7 & const & 0.001 & 100 & 100 & 1 & 1 \\
  & 8 & expo & 0.001 & 14000 & 100 & 1 & 1 \\
  & 9 & expo & 0.001 & 1000 & 100 & 1 & 1 \\
  & 10 & const & 0.004 & 14000 & 100 & 1 & 1 \\
  & 11 & const & 0.004 & 10 & 100 & 1 & 1 \\
  & 12 & const & 0.004 & 100 & 100 & 1 & 1 \\
  & 13 & const & 0.004 & 100 & 100 & 1 & 0 \\
  & 14 & expo & 0.004 & 14000 & 100 & 1 & 1 \\
  & 15 & expo & 0.004 & 1000 & 100 & 1 & 1 \\
  & 16 & const & 0.020 & 10 & 100 & 1 & 1 \\
  & 17 & const & 0.020 & 100 & 100 & 1 & 1 \\
  & 18 & const & 0.020 & 100 & 100 & 1 & 0 \\
  & 19 & expo & 0.020 & 14000 & 100 & 1 & 1 \\
  & 20 & expo & 0.020 & 1000 & 100 & 1 & 1 \\
  & 21 & const & 0.040 & 10 & 100 & 1 & 1 \\
  & 22 & const & 0.040 & 100 & 100 & 1 & 1 \\
  & 23 & const & 0.040 & 100 & 100 & 1 & 0 \\
  & 24 & expo & 0.040 & 14000 & 100 & 1 & 1 \\
  & 25 & expo & 0.040 & 1000 & 100 & 1 & 1 \\
\noalign{\smallskip}  
\hline  
\end{tabular} \\   \end{flushleft}
\label{semstab} 
\end{table}

  \begin{figure*}
  \includegraphics[width=18cm]{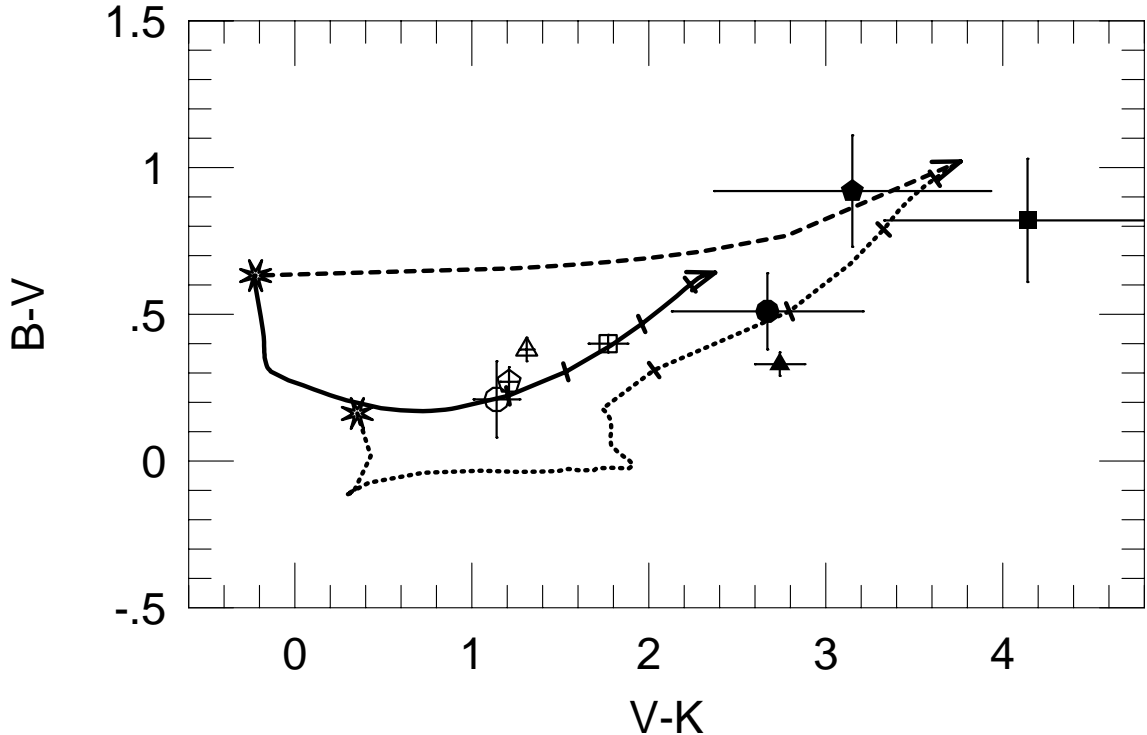}
  \caption[]{The predicted evolution of $B-V$ and $V-K^\prime$ of
of a star formation galaxy with an e-folding decay rate of 10$^9$ years.
Two scenarios (models no. 1:9 and 2:25 in table 6) with different metallicities, 
5\% solar (solid line) 
and twice solar (dotted line), are shown. A Salpeter mass function 
is assumed. The burst starts at the stellar symbol and ends after
14 Gyr at the arrows. The hatches along the evolutionary tracks
mark the evolution at 1, 2, 4 and 8 Gyr.  The hatched line indicates the 
theoretical 
position of a young burst
superposed on the old metal-rich population if we continuously vary the
relative contribution of the burst to the integrated light from 0 to 100\%. Data 
for 
the central 
burst (unfilled symbols) and and the halo (filled symbols) of
the target galaxies are shown. The galaxies are ESO 338-IG04 (triangle), 
350-IG38 
(square), 400-G43 (pentagon) and 480-IG12 (circle). 
Reddening corrections based on spectroscopy of the  
central region and assuming the column density of the dust to be proportional
to the luminosity density, have also been applied.}  
\label{bvvkshort}   \end{figure*}

  \begin{figure*}
  \includegraphics[width=18cm]{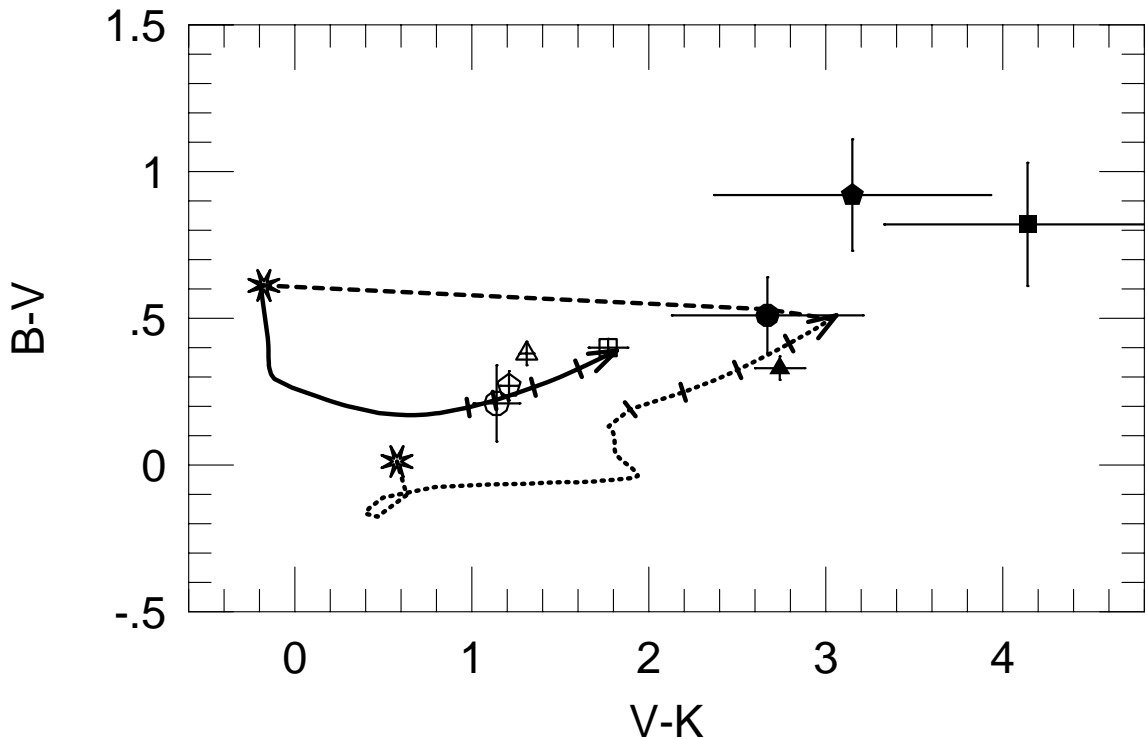}
  \caption[]{Similar to Fig. \ref{bvvkshort} with the difference that the decay 
rate of the star formation is 14 Gyr (models 1:8 and 1:24 in table 6).}  
\label{bvvklong}   \end{figure*}

  \begin{figure*}
  \includegraphics[width=18cm]{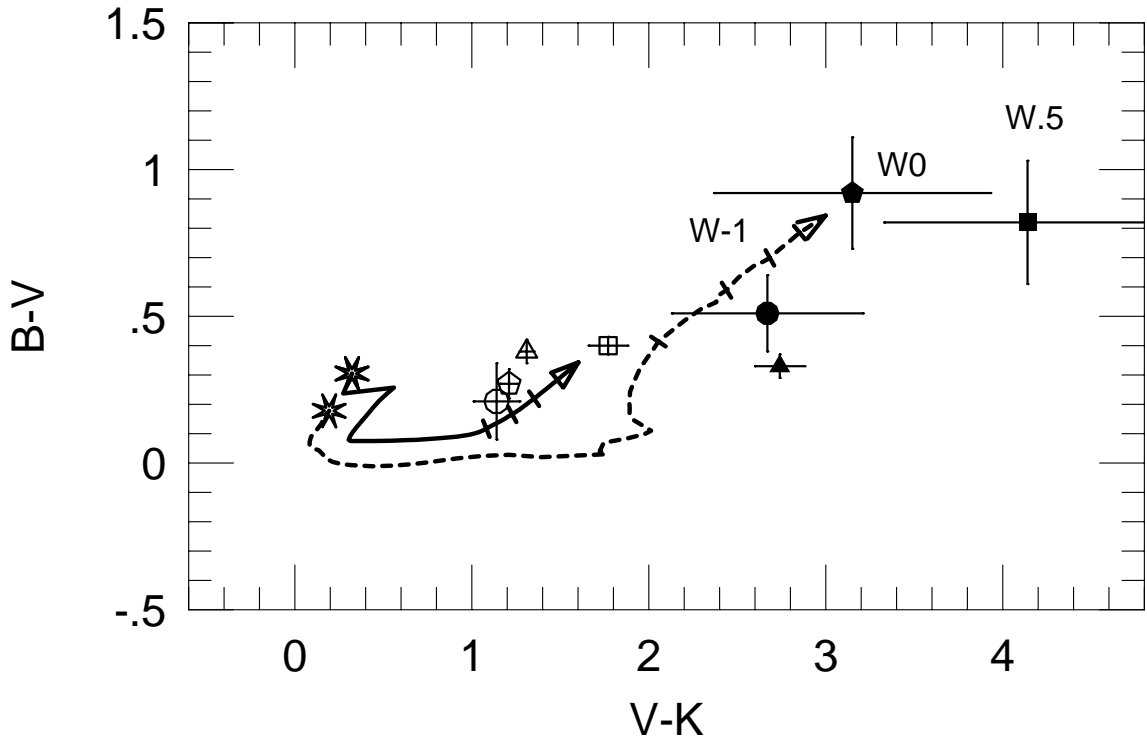}
  \caption[]{The diagram shows the predicted evolution from the PEGASE model 
(Fioc \& Rocca-Volmerange 
\cite{fioc}) in $B-V$ and $V-K^\prime$ of a star forming galaxy with constant 
star
formation rate and zero initial metallicity (full drawn line) and a constant 
star 
formation rate during the initial 10$^8$ years and an initial metallicity of 5\% 
solar
(Z=0.001; hatched line). The bursts start at the stellar symbols on the left
and end after 14 Gyr at the arrows. The final mass weighted stellar
metallicities are about 10\% solar and 1.2 times solar respectively. Also
indicated are the predicted colours from Worthey (\cite{worthey}) of an old
(12 Gyr) single-burst stellar population of three different metallicities,
[Fe/H]=$-1.0$, 0.0 and 0.5. These are indicated by the symbols W-1, W0 and W05
respectively. The observational data are displayed as in Fig.~\ref{bvvkshort}}  
\label{bvvkpegase}   \end{figure*}

  \begin{figure*}
  \includegraphics[width=18cm]{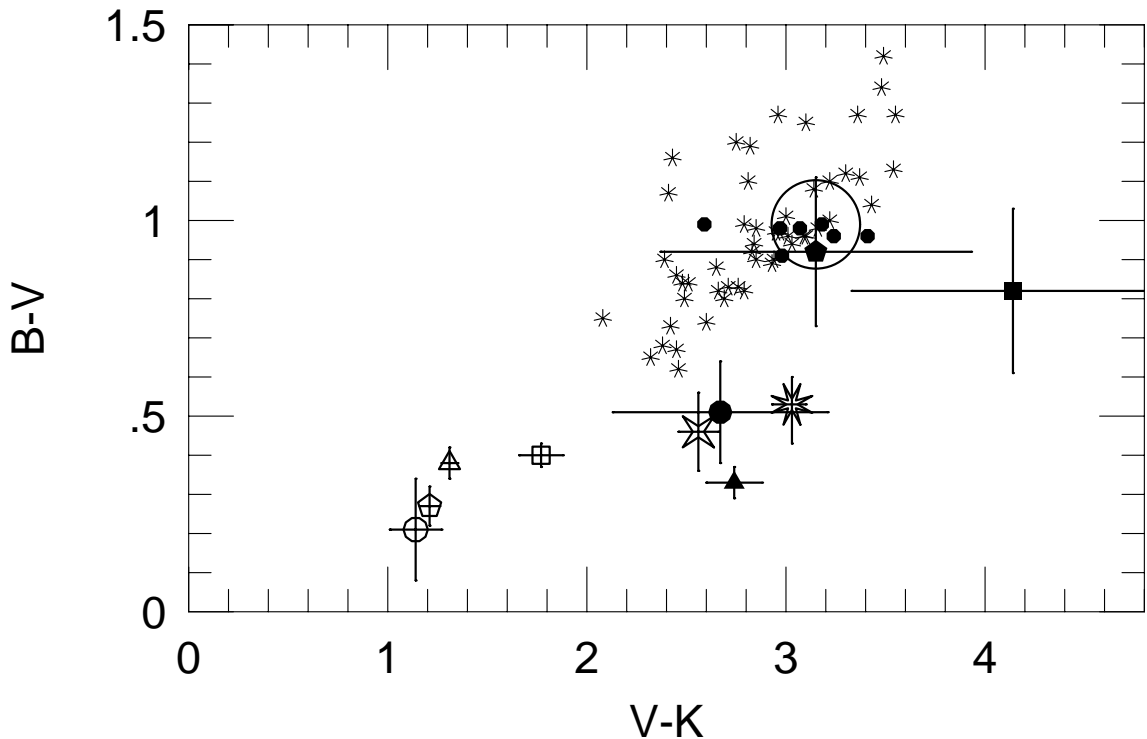}
  \caption[]{The observed colours of isolated E/S0 galaxies (Bergvall \& 
Johansson 
\cite{bergvall7}; filled dots), M31 globular clusters (Barmby et al.
\cite{barmby}; stars), the mean position of local ellipticals (Pahre
\cite{pahre}; large circle), the disk  of the luminous LSBG 0237-0159 (Bell et
al. \cite{bell}; big star), the low luminosity elliptical ESO 118-G34 (Sadler
et al. \cite{sadler}; diamond). Our BCG data are displayed as in
Fig.~\ref{bvvkshort}}    \label{bvvkcomp}   \end{figure*}

\begin{table*}
 \caption[ ]{{\bf Halo broadband colours and fits to the spectral evolutionary 
model.}
The table shows the integrated colours of the outermost region containing useful 
near-infrared data and the best model fit to the data. The model is based on two 
stellar components, one which is assumed to have a metallicity of 5\% solar 
(the first model of the two). The metallicities of the second component ranged 
from 5\% to twice solar. The allowed age range was 0 to 15 Gyr. A Salpeter IMF 
was assumed. The model numbers refer to the numbers in Table 6. The fits were 
made by mixing the spectral distributions of two models, one from 
each of the two groups in the table, giving them weights that minimized the
deviation from the observations. Among these, the mix that resulted in the best 
fit to the data was finally selected and presented below. The table shows the 
colours, total M$_V/L$ ratios (i.e. including dark baryonic matter) and masses 
of the two components and the mixture between the two. $\Delta _{Halo-Mix}$ is 
the difference between the observations
and the model predictions. The masses were calculated from the estimated halo 
luminosity based on the power law fit in Table 5. At the end of respective row 
$\sigma$ is the weighted mean error of the observed colours and the fits and 
$\sigma_{mp}$ is the weighted mean error of the best fit
when only metal-poor ($\leq$ 20\% solar) models were used.}  
\begin{flushleft}   
\begin{tabular} {lllllllllllllll}   
\noalign{\smallskip}  
\hline   
\noalign{\smallskip}
Galaxy & Component & Age & M$_B$ & B-V & B-R & V-I & V-K & J-H & H-K & M$_V/L$  
& 
Mass & $\sigma$ & $\sigma_{mp}$ \\
& & (yr) & & & & & & & & & (\sma) & mag. & mag. \\
\hline 
\noalign{\smallskip} 
338-04  & Halo                  &        & -17.70 &   0.31 &   0.69 &   0.70 &   
2.83 &   0.68 &   0.12 & \\
        & $\sigma_{obs}$        &        &   0.02 &   0.03 &   0.03 &   0.05 &   
0.24 &   0.36 &   0.35 & & & 0.017 \\
        & Model 1               & 4.5e8  & -16.24 &   0.17 &   0.13 &  -0.06 &   
0.67 &   0.25 &   0.15 &   0.16 &  5.0e7 \\
        & Model 24              & 7.5e9  & -17.36 &   0.39 &   0.85 &   0.88 &   
2.73 &   0.57 &   0.25 &   2.35 &  2.5e9 \\
        & Mix                   &        & -17.69 &   0.34 &   0.70 &   0.73 &   
2.50 &   0.56 &   0.25 &   1.85 &  2.6e9 \\
\medskip
        & $\Delta _{Halo-Mix}$ 	&        &  -0.01 &  -0.03 &  -0.01 &  -0.03 &   
0.33 &   0.12 &  -0.13 &  &  & 0.017 & 0.032 \\
338-04b & Halo                  &        &   -    &     -  &    -   &   0.62 &   
2.87 &   0.42 &   0.41 & \\
        & $\sigma_{obs}$        &        &   -    &     -  &    -   &   0.03 &   
0.28 &   0.17 &   0.28 & & & 0.02 \\
        & Model 9               & 5.0e5  & -16.95 &   0.63 &   0.00 &  -1.46 &  
-0.22 &   0.04 &   0.58 &   0.005 &  4.6e6 \\
        & Model 22              & 6.5e9  & -16.55 &   0.94 &   1.63 &   1.34 &   
3.55 &   0.64 &   0.28 &   7.39 &  6.3e9 \\
        & Mix                   &        & -17.52 &   0.77 &   0.96 &   0.63 &   
2.79 &   0.62 &   0.29 & \\
\medskip
        & $\Delta _{Halo-Mix}$ 	&        &   -    &     -  &    -   &   0.00 &   
0.08 &  -0.20 &   0.12 &  & & 0.01 & 0.13 \\
350-38  & Halo                  &        & -20.00 &   0.87 &   1.24 &    -   &   
4.16 &    -   &   0.82 & \\
        & $\sigma_{obs}$        &        &   0.03 &   0.26 &   0.12 &    -   &   
0.81 &    -   &   1.20 & & & 0.21 \\
        & Model 9               & 5.0e5   & -19.16 &   0.63 &   0.00 &  -1.46 &  
-0.22 &   0.04 &   0.58 &  0.005 &  3.5e7 \\
        & Model 23              & 14.5e9 & -19.33 &   1.02 &   1.76 &   1.46 &   
3.77 &   0.66 &   0.28 &  14.92 &  1.8e11 \\
        & Mix                   &        & -20.00 &   0.86 &   1.26 &   1.00 &   
3.28 &   0.65 &   0.28 &  9.35  &  1.8e11 \\
\medskip
        & $\Delta _{Halo-Mix}$  &        &   0.00 &   0.01 &  -0.02 &    -   &   
0.88 &  -0.65 &   0.54 & & & 0.09 & 0.22 \\
400-43  & Halo                  &        & -19.50 &   0.92 &   0.74 &    -   &   
3.15 &   0.95 &   0.71 & \\
        & $\sigma_{obs}$        &        &   0.03 &   0.19 &   0.19 &    -   &   
0.78 &   0.63 &   0.86 & & & 0.13 \\
        & Model 6               & 5.0e5   & -19.15 &   0.65 &  -0.03 &  -1.50 &  
-0.25 &   0.04 &   0.58 &   0.005 &  3.6e7 \\
        & Model 22              & 13.5e9 & -18.10 &   1.02 &   1.76 &   1.46 &   
3.76 &   0.66 &   0.28 &  13.59 &  5.2e10 \\
        & Mix                   &        & -19.50 &   0.76 &   0.80 &   0.44 &   
2.66 &   0.63 &   0.29 &   4.73 &  5.2e10 \\
\medskip
        & $\Delta _{Halo-Mix}$  &        &   0.00 &   0.15 &  -0.07 &    -   &   
0.48 &   0.31 &   0.42 & & & 0.14 & 0.31 \\
400-43b & Halo                  &        & -17.40 &    -   &   0.89 &    -   &   
{\it 2.88}$^1$ &  -0.17 &   0.62 & \\
        & $\sigma_{obs}$        &        &   0.05 &    -   &   0.09 &    -   &   
{\it 0.22}$^1$ &   0.26 &   0.31 & & & 0.12 \\
        & Model 4               & 2.5e9  & -17.28 &   0.41 &   0.78 &   0.70 &   
{\it 2.24}$^1$ &   0.41 &   0.13 &   1.76 &  1.8e9 \\
        & Model 22              & 10.5e9 & -14.97 &   1.00 &   1.73 &   1.43 &   
{\it 4.72}$^1$ &   0.65 &   0.28 &  10.76 &  2.2e9 \\
        & Mix                  	&        & -17.40 &   0.49 &   0.93 &   0.86 &   
{\it 2.96}$^1$ &   0.53 &   0.21 &   3.30 &  4.1e9 \\
\medskip
        & $\Delta _{Halo-Mix}$ 	&        &   0.00 &    -   &  -0.04 &    -   &    
   -0.08      &  -0.70 &   0.41 & & & 0.16 & 0.19 \\
480-12  & Halo                  &        & -19.30 &   0.51 &   0.79 &    -   &   
2.67 &   0.93 &  -0.38 & \\
        & $\sigma_{obs}$        &        &   0.03 &   0.13 &   0.09 &    -   &   
0.54 &   0.31 &   0.56 & & & 0.14 \\
        & Model 9               & 3.5e7  & -18.90 &   0.26 &  -0.04 &  -0.61 &   
0.05 &   0.12 &   0.29 &   0.03 & 9.8e7 \\
        & Model 23              & 14.5e9 & -18.01 &   1.02 &   1.76 &   1.46 &   
3.77 &   0.66 &   0.28 &  14.92 & 5.2e10 \\
        & Mix                   &        & -19.29 &   0.55 &   0.87 &   0.81 &   
2.99 &   0.64 &   0.28 &   7.05 & 5.2e10 \\
        & $\Delta _{Halo-Mix}$  &        &  -0.01 &  -0.04 &  -0.08 &    -   &  
-0.32 &   0.30 &  -0.66 &  &  & 0.11 & 0.32 \\
\noalign{\smallskip}  
\hline  
\end{tabular} \\ 
1) The data refer to B-K instead of V-K  
\end{flushleft}
 \end{table*}

\section{Ages and metallicities - model comparisons}

 We will now use the colours of the burst and halo regions 
 together with the spectral evolutionary models (SEMs) to
 constrain the age and metallicity of the stellar populations.
 The predicted colour evolution of the SEMs depend on
 the assumed initial mass function (IMF), metallicity and 
 a parameter describing how the SFR changes with time. 
 More sophisticated models also include metallicity 
 evolution and gas flows.
 There is still on going debate about the slope of the IMF 
 and its possible universality, but within the range 
 consistent with observations (e.g. Salpeter, Scalo, Kroupa
 et al, Miller-Scalo etc), varying the IMF has a relatively
 small impact on colours, except at the very earliest stages.
 This is because the emitted light will be dominated by
 the most luminous stars still alive. Sometimes, IMFs 
 truncated at low stellar masses have been proposed for 
 starburst regions. However, at low ages, low mass stars are
 insignificant contributors to the emitted light, whereas
 at high ages they are needed to get any light at all. 
 Hence varying the IMF, and its lower and upper mass limits,
 will affect the mass to light ratio, but have rather small 
 impact on the colour evolution of a stellar population. 
  On the other hand, metallicity and the temporal 
 behaviour of the SFR has a major impact on colours. Assuming that
 the SFR is constant in time leads to blue colours at high
 ages, as compared to a short burst. A correct treatment 
 of nebular emission is crucial at low ages, but less important
 a high ages. Hence, in order to restrict the parameter space, 
 we have chosen to use only models with a ``normal'' Salpeter 
 IMF, but instead let the metallicity and star fomation
 history vary and also and the density, covering factor 
 and filling factors of the ionised gas (with a single,
 observationally known, metallicity). In Table \ref{semstab}, 
 we give a list of the used parameters in the SEMs.

\subsection{Location in the $B-V$ vs $V-K$ diagram}

Before entering into the detailed modelling of the halo colors, 
we will have a look at the location of the galaxies in the $B-V$ 
vs $V-K$ two colour diagram.
In Fig.~\ref{bvvkshort}, \ref{bvvklong},
\ref{bvvkpegase} and \ref{bvvkcomp} we try to show, as  simple as possible,
the main   restrictions we can impose on the stellar populations in the   
galaxies. The $V-K/B-V$ diagrams show the predicted colour    evolution from 0
to 14 Gyr of a star forming galaxy with different star formation  histories
and metallicities. For comparison we show the results from three different
SEMs from three different research groups.  

In Fig.~\ref{bvvkshort} we show the predicted evolution of an exponentially  
decaying star formation history with timescale of 1 Gyr at two different 
metallicities, 5\% solar and twice solar (Zackrisson et al.  \cite{erik}). 
Fig.~\ref{bvvklong} shows the same data with the difference that the star 
formation decay rate is now 14 Gyr. Hence these represent 4 extreme cases,
and the truth may well lie in between. An instant burst model would differ 
from the $\tau = 1$ Gyr model, only for ages less than a few Gyrs. However,
real galaxies can hardly be truly instantaneous on a global scale, but a 
time scale of 1 Gyr is probably more realistic. In Fig.~\ref{bvvkshort} 
($\tau = 1$Gyr) one sees that the the colours of the starburst regions seem 
to fall along the metal-poor track, whereas , surprisingly (given the
observed nebular metallicities),  the halo colours lie close to the old 
ages of the metal-rich track. The same pattern is seen Fig.~\ref{bvvklong} 
($\tau = 14$ Gyr), but now the two reddest halos are redder than even the
oldest points in the metal-rich model. Hence, this first simple look at
the halo colours suggests that the halos are old,  metal-rich and 
formed on relatively short timescales. 

In Fig. \ref{bvvkpegase} we use the predictions from the PEGASE2 model (Fioc \&
Rocca-Volmerange \cite{fioc}), in which the recycling of metals in a closed
box scenario have been taken into account. Two tracks are displayed - one
shows the evolution of a stellar population with zero initial metallicity
assuming the star formation rate is kept constant. The other track shows the
evolution starting from an initial metallicity of approximately 5\% solar 
(Z=0.001)
and then evolving with an exponentially declining SFR and a decay rate of 1
Gyr, as in Fig.~\ref{bvvkshort}. In this diagram we also display the
predictions of colours of an old stellar population of 3 different
metallicities ([Fe/H]$=-1.0, 0.0$ and 0.5) from Wortheys (\cite{worthey}) 
models.
On top of this we display the colours of the central region and the halo of
our BCGs. Again, it appears that the reddest halos are old and quite metal-rich.

As can be seen, the colours of the central starburst agree very well with the 
predictions from the metal-poor model. An important fact to remember is that all 
4 BCGs show Wolf-Rayet features in their spectra and have strong Balmer emission 
lines and thus must contain a considerable fraction of very young stars, as
is also indicated by the photometry in the UV/blue region and UV spectra
obtained with IUE. What may seem somewhat  unexpected perhaps is
that the luminosity weighted ages of the burst populations, as found from the
positions in the diagrams, are quite high, as if the star formation has been
going on continuously for more than 1 Gyr. This cannot be correct because it is 
not consistent with the high equivalent widths of the observed emission lines. 
The equivalent width of \ha is several hundred \AA ~and the luminosities 
correspond 
to star formation rates of several \sm yr$^{-1}$.  The available gas budget 
of a few times 10$^8$\sm would be thus consumed in about 10$^8$ yr (or even 
faster if we have a decaying SFR) unless the IMF is peculiar. The explanation of 
the 
colours is probably that there is a burst occurring in an old host. 
In Fig. \ref{bvvkshort} and \ref{bvvklong}
we show the effect on the colours of the transition from a passive old
population to a young burst. Depending on the age and the relative mass of the
burst it may take any position within the region defined by the connecting
lines between the young and old population. Although the optical region
may be little affected by the mixing of the two population we may notice it in
the near-IR, in $V-K$.  A mixed population 
with ~10\% of the $V$-band luminosity from a population with the same color
as the ones observed in the halos and a 90\% from a young (~$10^7$ yr) model 
population gives composite colors which agrees well with those of the central 
regions. Thus it is perfectly proper to assume that the ages of the burst 
population 
are quite low, i.e. $\leq$ 10 Myr.

\subsection{Best fitting models to the halo colours}

Let us now look a bit more detailed at the spectral energy distribution of
the halos and the best fitting SEM predictions from the Zackrisson  et al.
(\cite{erik}) model. For simplicity and considering the large uncertainty 
in the near-IR colours of the halo, a few parameters of the SEMs were kept 
fixed. Thus we assumed a Salpeter IMF, a mass range between 0.08 and 120 
\sma and a constant metallicity. 

We mixed two models where the metallicity in both gas and stars 
in one of the models was fixed to 5\% of the solar (set no. 1 in Table 
\ref{semstab}). We think this is a reasonable guess of the metallicity 
of the gas used in the burst. It is also at the lower envelope of the 
metallicity of the (recently enriched) ISM of gas-rich LSB galaxies 
which, as we will argue below, are attractive merger candidates. 
Hence this metal-poor component would allow us to simulate contamination
to the halo colours from young stars.
The metallicity of the second 
component, which we may associate with the halo population, was allowed to vary 
between 5\% solar and twice solar (set no. 2 in Table 
\ref{semstab}). The star formation rate was either assumed to 
occur in a burst of constant SFR for $t\le \tau$ or in an exponentially 
declining 
mode, SFR $\propto$ e$^{-{t\over\tau}}$, where $t$ is the age of the burst and 
$\tau$ 
is the duration of the burst or the SFR decay timescale. Table 6 lists the 
parameters of the models we included in the comparison. The optimized mixture of 
the two sets of models were fitted to the observed fluxes that were given 
weights 
in proportion to the square of the inverse of the mean error.

Table 7 lists the models that, when optimally mixed, gave the best fits to 
the observed broadband fluxes. The table displays predicted colours and the 
deviations from the observations. Also inluded are the weighted mean errors 
of the observations and the mean error of the fit. The errors refer to the 
colours and  are based on the weighted errors of the fits to the fluxes. 
Although the errors are quite large (we are investigating regions with very
low surface brightneses), the best fits are with a few exceptions obtained 
within about 1 $\sigma$ of the observational errors. It is noteworthy
that we find several cases of large deviations with respect to the observed 
$V-K$ colours, in the sense that observations are redder than the models.
Hence, increasing the weight of the near IR data would lead to higher
ages and metallicities.
The extremely young ages derived for the metal-poor
component in 2-3 cases is of course unrealistic and should be regarded as 
indications of low age only. The important result is that, even as we include 
all available photometry in the comparisons, we get consistency with the 
conclusions from the previous discussions - the halo components of {\it all 
galaxies} 
have colours that best fit with models of an {\it old metal-rich stellar 
population}.

For comparison, we have also included in the table the mean error of the
best fit to a model including only metal-poor ($<$ 20\% solar) stellar 
populations.
These fits are significanlty worse except for one case, ESO 400-G43b. In
this case the fit is equally good if we assume a low metallicity throughout.
It is interesting to note that this is also the only galaxy in our sample
that has a luminosity profile that fits well with an exponential disk.
It thus has properties similar to LSB galaxies that are thought to have
had a rather constant and low star formation activity over a long time
and are probably not sitting in massive halos.
The other cases typically deviate with 1-3$\sigma$ from the observations
and are thus 2-3 times worse when we force the halo to have a low metallicity.
While the errors are large, and the sample is small, the comparison
with the SEMs suggest that the halos are more metal-rich than inferred
from the nebular gas.

The predicted photometric masses of the halos, based on the halo luminosities 
from Table 5, range from $3 \times 10^9$ to $2 \times 10^{11}$ \sma and should 
be 
regarded 
as upper limits. Had we chosen the Scalo (\cite{scalo}) IMF instead of the 
Salpeter IMF,  the predicted masses would be reduced with a factor $\approx$ 2.
Moreover, had we used an exponential luminosity profile in estimating the 
the total halo luminosities, the mass estimates would go down with up to
a factor of 10.

\section{Discussion -- Evolutionary scenarios}

Now let us have a look at the halo colours and discuss the properties of the 
host 
of
the starbursts. Let us first recall the most popular alternative star
formation scenarios discussed in this context: 1) What we observe is the first
star formation epoch in the history of the galaxy. The galaxy is young. 2)
This is one of several intermittent bursts taking place in the galaxy. The
bursts are caused by infall of cooled gas from the halo, processed in a
previous burst. 3) Interaction with a neighbor,
causing gas from the outskirts to fall towards the centre, thereby igniting a
burst 4) Merger between galaxies or one galaxy (or a
few galaxies) and intergalactic gas clouds. We will now discuss these
alternatives one by one.

\subsection{Young galaxy} 
The fact that the chemical abundances are low and
homogeneous agree with a young galaxy but the colours and the morphology of the 
halo
population do not. We have shown that the light from the halo is dominated
by stars. Thus the galaxy has to be at least as old as the time it takes for
the stars to relax into a regular structure. This is at least a few crossing
times or a few hundred Myr. The colours of the halo populations are much redder 
than 
that of metal-poor gas-rich low surface brightness galaxies (BCG halo: 
$B-J=2.6\pm0.2$
mag. as compared to LSB: $B-J=1.5\pm0.2$ mag.; Bergvall et al. \cite{bergvall6}) 
and
do not at all agree with a young metal-poor population. The agreement is fine 
however, 
with a {\it metal-rich
stellar population of an intermediate-high age ($\approx$ 2-15 Gyr)}. This is 
evident  
from the
figures where the model predictions are displayed and is consistent  with all
three model codes we compare with. Both ESO 350-IG38 and 400-G43 appear to have
old halos while ESO 338-IG04 and 480-IG12 may have halo ages of between 1 Gyr 
and a Hubble age, depending on the contribution 
of young stars to the halo light.

In Fig. \ref{bvvkcomp} we compare with different samples of observational data 
from the literature. This includes a set of globular clusters in M31 (Barmby
et al. \cite{barmby}) a sample of seemingly isolated E/SO galaxies (no bright
neighbors in projection; Bergvall \& Johansson \cite{bergvall7}) and another
sample of mostly cluster ellipticals (Pahre \cite{pahre}), the distribution of
which is indicated by a large circle in the diagram. The range of metallicities 
of the globulars is -2.0$\leq$[Fe/H]$\leq$0, with mean metallicities slightly 
below [Fe/H]=-1.0. This is close to the metallicity of the burst population of 
our target
galaxies. Yet they deviate strongly from the colours of the halos. This is in 
disagreement with Doublier et al. (\cite{doublier2}). They however studied the
global colours and their sample was selected differently. Admittedly
the error bars of the halo colours are quite large but taken together we find
no other reasonable explanation than high age and high metallicity (taking the
risk of generalizing the properties of the halos of the four galaxies). We
have no reason to suspect that we have serious calibration problems and a 
comparison 
with the few other photometric data available is reassuring. E.g our photometry 
of 
the galaxy with the most extraordinary colours, ESO\,350-IG38, agree remarkably 
well 
with those of Vader et al (\cite{vader}). They obtain $M_B=-20.2$, $K=12.08$, 
$J-H=0.62$, 
$H-K=0.62$ ($D=9.3\arcsec$) We obtain  $M_B=-20.0$, $K=12.06$, $J-H=0.67$, 
$H-K=0.62$. 

\subsubsection{Gas and dust emission}

In Sect. 6 we concluded that the influence of gas and dust on the colours is 
small. How 
much and in which direction would the colours be affected if we relax these
constraints? We have discussed some constraints on the \ha emission that cover 
the 
region including the infrared data of ESO 350-IG38, 338-IG04 and 400-G43. In 
these two
cases we can rather safely say that the contribution from nebular emission to
the halo light is negligible. But in the other cases we cannot exclude that
the contribution from nebular emission increases as we reach the red regions.
It is therefore motivated to have a look at how nebular emission would affect
the colours. We can derive the colours of a pure HII region (i.e assuming no
stars in the halo so that 100\% of the flux is due to the ionised gas)
assuming that the metallicity is the same as that of the ionised gas in the
centre. The colours we obtain are $B-V\le$0.74 and $V-K\le -0.17$, i.e.
significantly bluer than observed. Thus, if anything, the colours of the
stellar population would become redder if we correct for the contribution from
emission from gas ionised by the central source. With the assumptions made in 
Sect. 6 the correction in $B-V$ would be marginal while in $V-K$ it would amount 
to 
0.1-0.2 magnitudes.

Could the emission be due to warm dust? We may constrain the properties of such 
a component from the $K$ band photometry combined with ISOCAM 4-14$\mu$ data of 
ESO 350-IG38 from the ISO mission (Bergvall et al., in preparation). The total 
fluxes we
obtain from the ISOCAM data are 4.0$\mu$: 23.6$\pm$0.3 mJy,
6.7$\mu$: 70.6$\pm$0.4 mJy, 9.6$\mu$: 151$\pm$1 mJy, 14.3$\mu$: 589$\pm$3 mJy. 
If
we assume an emissivity of $\epsilon \propto$ B$_\nu \lambda^{-\beta}$, with
1.4$\le \beta \le$2.0 (Lisenfeld et al. \cite{lisenfeld}) we obtain a
temperature range of 100-160 K based on these global ISO data. If the heating
is caused by the central burst, 160K is the maximum temperature we can accept
in the halo. The total K band flux density is 135 mJy. 10\% of this emission
originates from the region outside 5 kpc that has a clear red excess. The excess 
in $K^\prime$ relative to a metal-poor stellar population is $\sim$ 1.4$^m$, or
roughly 80\%. If we therefore assume that at least 50\% of the $K^\prime$ 
emission,
i.e. $\sim$ 7 mJy, comes from warm dust we can calculate a {\it minimum}
temperature of the dust from the 2$\mu$ and the 4$\mu$ fluxes. The 4$\mu$
emission outside 5 kpc contributes with $\sim$ 5\% of the total emission in
this frequency band. With these data, the temperature we derive is around
3500K, which is impossible if the dust is heated by the central burst and
unreasonable if the dust is heated in situ. Warm circumstellar dust with a
temperature of 500K would contribute with 0.1\% of the K flux from the halo if
we scale it with the 4$\mu$ emission. We conclude that the contribution from
dust emission to the halo $K^\prime$ luminosity is insignificant.

Since the dust extinction to the amount that would be 
needed for the data to be consistent with a young age, i.e. A$_B \sim$ 1 
mag., probably can be excluded, the only alternative remaining, if we stick
to low metallicities, is that the IMF is odd. But, if so, the only option is a
very low upper mass limit, a highly improbable situation if we can trust the
theoretical models of star formation. The youth hypothesis therefore
can be dismissed. 

\subsection{Recurrent bursts} 
We have shown in a previous paper (\"Ostlin et al. 
\cite{goran1}) that at least ESO 338-IG04 has been involved in 
previous bursts. But if the gas in the present burst originates from expelled 
gas from a previous burst, the metallicity of the gas must be higher than or 
comparable to that of the host, not an order of magnitude lower. Following 
the conclusions from the previous discussion, this is not the case. We can
therefore conclude that recurrent bursts, controlled by the cyclic expulsion 
and later accretion of the ISM, cannot be the explanation of the starburst
seen in our BCGs. Previous bursts induced by infall of fresh material, either
from gas-rich dwarfs or in the form of intergalactic HI clouds, are however not 
excluded.

\subsection{Tidally induced burst} 
Gas rich galaxies of low to intermediate 
masses have small abundance gradients. Thus if the starburst is caused by
interaction with a neighboring galaxy we still would have to explain why the
metallicity of the gas is lower than that of the old stars. Moreover the
statistics give no observational support of strong starbursts caused by
interaction between low mass galaxies. Only gas-rich LSB galaxies have the
required properties of the metal-poor progenitor and it seems to be very
difficult to ignite a starburst in such a galaxy (Mihos et al. \cite{mihos}). 

\subsection{Merger} 
A merger between two galaxies of different metallicities, 
one gas-rich and one gas-poor, or infall of intergalactic clouds, appear to
be the most plausible explanation of the observational results. We have argued
in previous papers (Bergvall et al. \cite{bergvall5}, \cite{bergvall6}) that
gas-rich LSB galaxies, and perhaps only this type of galaxy and not normal
irregulars, have the required properties of the metal-poor component. 
One of the arguments was based on the location of these galaxies in the 
metallicity--luminosity diagram (Fig. \ref{mz}). If one assumes the 
progenitor to be a dI, the offset of more than 3 magnitudes in luminosity 
from the dI relation would require an excessive neutral hydrogen gas 
mass.  LSBGs on the other hand already lie offset from the dI relation,
meaning that they need to brighten less, and in addition, the gas mass 
fractions are much higher than for dIs. Note that also IZw18 and SBS\,0335-052
show the same  offset from the  $M_B - Z$ relation in Fig. \ref{mz}. 

The old halo component could be a dwarf elliptical, a massive ellipsoidal 
disk or a similar type of galaxy.
If the last interpretation is correct, it would indicate that there should be 
cases in which we would find low luminosity early type galaxies hosting HI
clouds or gas-rich galaxies in a state prior to a burst. Sadler et al.
(\cite{sadler}) observed four E/S0 and S0 galaxies with these properties.
Their sample galaxies have luminosities ranging from $M_B=-17.0$ to $-19.1$, 
i.e. similar to 
the maximum luminosities of our host galaxies, and contain HI
with masses in the range 10$^8$ to 10$^9$ \sma. They found that the HI is
strongly concentrated towards the central region, differing from what is found
in normal early type galaxies. In two of the cases the distribution of the HI
resembles warped disks like in ESO\,338-IG04 and ESO\,480-IG12. Similar results
were obtained by Lake et al. (\cite{lake}) for another sample of low
luminosity early type galaxies. James and Mobasher (\cite{james2}) find star
formation activity most frequent in ellipticals situated in regions of low
density but not completely isolated. Two of the galaxies in the Sadler et al.
sample show clumpy H{\sc ii} emission in the centre with an \ha luminosity about 
a
few percent of what we observe in our BCGs. Interestingly enough, the
optical/near-IR colours are similar to the halos of our disk-like systems,
e.g. their ESO 118-G34 has the colours $B-V=0.46$ and $V-K=2.56$, similar to the
halo of ESO 480-IG12 ($B-V=0.51$, $V-K=2.67$).  It is also interesting 
that the halo colours of ESO\,338-IG04 and 480-IG12  agree with
the disk colours of a giant LSB galaxy observed by Bell et al. (\cite{bell}),
believed to have supersolar metallicities from their comparison with SEMs. The
positions of these galaxies have been indicated in Fig. \ref{bvvkcomp}.

\subsection{Host galaxies and dark matter}

It seems that the best candidate of the starburst host galaxy in our BCGs
is a galaxy of early morphological type.
We know that dEs have the highest fraction of dark matter of all galaxy types 
and 
that the DM density is extremely high, approximately 1000 times higher than
in large ellipticals. Here we have the preconditions needed for creating
dynamical instabilities in an infalling LSB disk, followed by inflows and a
starburst. But there is also another alternative. Massive ellipticals may form 
in the gravitational wells of large amplitude dark matter
fluctuations. In the early formation phase they consequently were much more 
dark matter dominated than they are today. In sparse regions the buildup of the 
stellar
component of the galaxies would proceed with a lower tempo and maybe there
we could find DM dominated {\it massive} Es that occasionally would be involved 
in
mergers. A massive DM halo would be more efficient in retaining metals
produced in a burst than a low mass one, thus explaining the red colours.
Support for mergers or infall taking place in isolated
early type galaxies of $L^*$ luminosities ($M_B\sim -20$) comes from e.g.
Colbert et al. (\cite{colbert}). They found evidence of recent mergers in
about 40\% of the cases and in higher proportions than found in richer groups.

Early type galaxies follow a colour-magnitude relationship that may be used to 
estimate the predicted luminosity of the host, provided it is a normal
early-type galaxy. We can for instance use the relationship based on cluster
ellipticals or compare to $V-K$ colours derived from models including effects of
infall of gas and SN generated outflows (e.g. Gibson \cite{gibson}; Ferreras
\& Silk \cite{ferreras}; Fioc \& Rocca-Volmerange \cite{fioc}). The luminosity
we would expect for a galaxy with colours similar to the halo colours of our
BCG sample is much higher than what we obtain from the photometry. There are
at least two possible ways to explain this difference. One is related to the
mechanism behind the metallicity-colour-luminosity relationship. This
relationship is mostly thought to be caused by the mass dependency of loss of
metals through stellar winds. According to the models by Mc Low and Ferrara
(\cite{maclow}) and Ferreras and Silk (\cite{ferreras}) mass loss from galaxies 
during a starburst phase is strongly dependent on the mass of the galaxy
and dwarfs may, depending on the morphology, lose most of the
metals produced in the starburst to the environment. If galaxy formation is
biased towards high density regions, it would mean that the field ellipticals
contain less baryons per dark matter mass unit, i.e. there would be an
anticorrelation between mass-to-luminosity and mean density of the
environment. The critical luminosity for mass loss would therefore decrease
and we might find low luminosity, metal-rich galaxies in sparse regions. From
time to time a gas-rich galaxy would fall into this potential well and start a
starburst similar to what we would expect to see at high redshifts.
It is interesting to note that the fit to the Sersic equation of the halo 
luminosity profile that was described in Sect. 4.5, resulted in very high 
values of 
the n parameter. Although individually these are quite uncertain values, these
data collectively also indicate that the mass of the host is large.
If ther merger hypothesis is valid, further support for massive halos comes
from n-body simulations of mergers. Dubinski et al. (\cite{dubinski}) found that
tidal tails do not develope in disk galaxies where the mass is stongly
dominated by the halo. Although we see some morphological signatures of the 
aftermaths of mergers, we do not see tails.

\subsection{A peculiar IMF?}
One may also consider the possibility of an abnormal IMF, which drastically 
could change the interpretation of the colours. More and more evidence is
accumulating however in the direction of a universal IMF. While local
variations may occur, the global properties closely agree with the classical
Salpeter IMF (Scalo \cite{scalo}). On the other hand there are strong biases
in IMF studies since they normally deal with star forming regions that are
luminous and the star formation efficiency is high. The low surface brightness
makes it problematic to investigate star formation regions in which the mean
density is low, as in e.g. low surface brightness galaxies. Dynamical
estimates of the mass and the \ml ~ratio are more reliable but also have their
limitations because of the uncertainties in inclination, contribution from
dark matter and due to distorsions in the velocity field, both in the central
area and in the HI distribution. A possible contribution from population III
stars with peculiar IMF cannot be excluded as an exotic candidate to explain 
the red excess.

\section{Conclusions}

We present the result of an optical/near-IR photometric/spectroscopic 
investigation of four luminous blue compact galaxies and their companions. The
chemical abundances of the BCG starbursts are low, typically 10\% of the solar 
values. 
The star formation is intense, corresponding to gas consumption timescales of 
the
order of 100 Myr only. We show that the main body is embedded in a huge halo
of ionised gas, possibly ionised by the central burst. The galaxies thus have 
properties we normally associate with young galaxies and one of our targets, 
ESO 400-G43, is in the list of  young galaxy candidates. We conclude however 
that all galaxies discussed here  contain a significant population of old stars.

We demonstrate that is is possible to disentangle a separate 'halo population' 
from the burst population. This very red component has colours and structural
properties typical of elliptical galaxies with metallicities considerably in
excess of that of those derived from the starburst H{\sc ii}-regions. From this 
we conclude that the starburst is not an internal affair but is caused by a 
galaxy merger or infall of gas from the intergalactic space. 

In the optical region the galaxy type that dominates the light is typical of 
gas-rich low surface brightness galaxies. The luminosity of the halo, as derived
from the empirical metallicity-luminosity relationship of normal galaxies, is 
much in excess of what is observed. A possible scenario is that we
witness mergers between gas-rich LSB galaxies or massive HI clouds and massive
and metal-rich, but extraordinary faint ellipticals. Such galaxies may exist
in low density regions as a consequence of a different distribution of baryons
and dark matter (the classical bias), causing the field ellipticals to contain
relatively more dark matter than the cluster ellipticals. Alternatively the
host may have a very peculiar IMF. 

The deep dark matter potential field of the Es may be necessary conditions for 
the burst to start. Our findings may be typical of luminous blue compact
galaxies in general and may prove to be important in the early formation of
galaxies. Similar starbursts occurring in more luminous ellipticals may be rare
because such galaxies have experienced sufficiently many starbursts to build
up an extended hot corona of ionised gas that will shield the galaxy from
infall of neutral gas by thermal conduction.

  \begin{acknowledgements} We thank Saga Dagnesj\"o, Steven J\"ors\"ater, Kjell 
Olofsson, Jari R\"onnback and Erik Zackrisson for their contributions to this
work and for stimulating discussions. This work was   partly supported by the
Swedish Natural Science Research Council.   \end{acknowledgements}

  \end{document}